\documentclass[review]{elsarticle}

\usepackage{
amsmath, amssymb}
\usepackage[strict]{changepage}
\usepackage{xcolor}
\usepackage[inline]{enumitem}
\renewcommand{\(}{\begin{equation}}
\renewcommand{\)}{\end{equation}}
\newcommand{\bea}{\begin{eqnarray}}
\newcommand{\eea}{\end{eqnarray}}

\newcommand{\beq}{\begin{equation}}
\newcommand{\eeq}{\end{equation}}
\renewcommand{\(}{\begin{equation}}
\renewcommand{\)}{\end{equation}}

\newcommand{\Mov}[1]{{\color{black
}{#1}}}

\begin{document}

\begin{frontmatter}

\title{On the change of old neutron star masses with galactocentric distance}

\author[ad1,ad2,ad3]{Antonino Del Popolo}
\ead{adelpopolo@oact.inaf.it}
\address[ad1]{Dipartimento di Fisica e Astronomia, University Of Catania, Viale Andrea Doria 6, 95125 Catania, Italy}
\address[ad2]{INFN sezione di Catania, Via S. Sofia 64, I-95123 Catania, Italy}
\address[ad3]{Institute of Astronomy, Russian Academy of Sciences, 119017, Pyatnitskaya str., 48 , Moscow}

\author[add1]{Maksym~{Deliyergiyev}}
\ead{maksym.deliyergiyev@ujk.edu.pl}
\address[add1]{Institute of Physics, Jan Kochanowski University, PL-25406 Kielce, Poland}

\author[adm1,adm2]{Morgan Le Delliou\corref{corauth}}
\cortext[corauth]{Corresponding author}
\ead{delliou@lzu.edu.cn,Morgan.LeDelliou.IFT@gmail.com}
  

\address[adm1]{Institute of Theoretical Physics, School of Physical Science and Technology, Lanzhou University, No.222, South Tianshui Road, Lanzhou, Gansu, 730000, Peoples Republic of China}
\address[adm2]{Instituto de Astrof\'isica e Ci\^encias do Espa\c co, Universidade de Lisboa, Faculdade de Ciˆencias, Ed. C8, Campo Grande, 1769-016 Lisboa, Portugal}

\author[adl1,adl2,adl3,adl4]{Laura~{Tolos}}
\ead{tolos@th.physik.uni-frankfurt.de}
\address[adl1]{Institut f\"{u}r Theoretische Physik, Goethe Universit\"{a}t Frankfurt, Max-von-Laue-Stra\ss{}e 1, 60438 Frankfurt, Germany}
\address[adl2]{Frankfurt Institute for Advanced Studies, Goethe Universit\"{a}t Frankfurt, Ruth-Moufang-Str.1, 60438 Frankfurt am Main, Germany}
\address[adl3]{Institute of Space Sciences (ICE, CSIC), Campus UAB, Carrer de Can Magrans, 08193, Barcelona, Spain}
\address[adl4]{Institut d'Estudis Espacials de Catalunya (IEEC), 08034 Barcelona, Spain}

\author[adf]{Fiorella~{Burgio}}
\ead{fiorella.burgio@ct.infn.it}
\address[adf]{INFN Sezione di Catania, Via S. Sofia 64, I-95123 Catania, Italy}

\begin{abstract}
We show that the pulsar mass depends on the environment, and that it decreases going towards the center of the Milky Way. This is due to two combined effects, the capture and accumulation of self-interacting, non-annihilating dark matter by pulsars, and 
the increase of the dark matter density going towards the galactic center. We show that mass decrease depends both on the density profile of dark matter, steeper profiles producing a faster and larger decrease of the pulsar mass, and on the strength of self-interaction. 
Once future observations will provide the pulsar mass in a dark matter rich environment, close to the galactic center, the present result will be able to put constraints on the characteristics of our Galaxy halo dark matter profile, on the nature of dark matter, namely on its annihilating or non-annihilating nature, on its strength of self-interaction, and on the particle mass.
\end{abstract}


\end{frontmatter}


\section{Introduction}
Dark matter (DM) is a key ingredient for models
that try to explain 
cosmological structure formation without modifying gravity. Although 
the gravitational effects of DM are well documented \citep{Betoule:2014frx,Ade:2013zuv},
direct detection of particles for this dominant matter component continues to elude proofs: in accelerators or in nuclear recoil experiments
\citep{Chatrchyan:2012me,ATLAS:2012ky,Agnese:2014aze,Angloher:2011uu,Felizardo:2011uw,Klasen:2015uma,Akerib:2013tjd,Ahmed:2010wy,Bernabei:2010mq,Aalseth:2010vx,Aprile:2012nq}, indirect WIMP annihilation searches \citep{Conrad:2014tla},
in DM stars \citep{Dai:2009ik,Kouvaris:2015rea} or in some other
indirect quests as illustrated in Refs.~\citep{Bertolami:2007zm,LeDelliou:2007am,Bertolami:2008rz,Bertolami:2007tq,Bertolami:2012yp,Abdalla:2007rd,Abdalla:2009mt,Delliou:2014awa}.

In this context, different testing avenues of possible DM effects
are welcome, such as in pulsars, i.e. rotating neutron stars (NSs), which provide the advantage of 
extreme densities and can accrete DM, 
thus straining 
the saturated neutron gas. The amount of DM acquired by a NS follows the Tolman-Oppenheimer-Volkoff (TOV)
equation
 \citep{Tolman:1939jz,Oppenheimer:1939ne},
as in e.g. \citep{Tolos2015}. 
Moreover, the effect of DM on NSs can directly lead to bounds for the masses of the different DM candidates
\citep{Goldman:1989nd,Kouvaris:2011fi}. 
 For example, a recently discovered class of radio transients, such as fast radio bursts \cite{Keane:2012yh, Thornton:2013iua, Burke-Spolaor:2014rqa, Spitler:2014fla}, can be caused by the collapsing NSs with accreted enough DM near GC \cite{Fuller:2014rza}.

Self-annihilating DM can also produce characteristic effects on NS \citep{Kouvaris2008,Bertone2008,Kouvaris2010,McCullough:2010ai,deLavallaz2010,PerezGarcia:2011hh,PerezGarcia:2010ap}.
In particular, WIMPs annihilation in DM cores should produce temperature and luminosity changes, through heat, of old stars \citep{Kouvaris2008,Bertone2008,Kouvaris2010,deLavallaz2010}. 
However, those changes 
are 
difficult to 
detect \citep{Kouvaris2008,Sandin2009}. 

Apart from WIMPs, DM could be made of asymmetric dark matter (ADM). In that case,  the origin of the present DM abundance is similar to visible matter \citep{Petraki2013}\footnote{Mirror matter is a peculiar case of ADM.}. As ADM does not annihilate, its collapse and thermalisation can give rise to extremely compact objects and to changes in the mass-radius ($M-R$) relation. The comparison of the $M-R$ relations coming from usual NSs and from NSs containing DM  could yield constraints on DM  and on the NS equation of state (EoS) \citep{Ciarcelluti2010}. Moreover, above a critical value of accumulated DM \citep{Kouvaris2013}, DM could become self-gravitating, form a mini black-hole, and constrain the DM particle cross section  and mass \citep{Bertone2008}.

Several authors realized in the last years that admixing DM with NS matter,
coupled only through gravity, has similar results as the presence of exotic matter \citep{Sandin2009}, allowing to explain very compact NSs  \citep{Ciarcelluti:2010ji},
 or very massive pulsars ($2M_{\odot}$, e.g., PSR J1614-2230 with $M=1.97 \pm 0.04 M_\odot$ \citep{demorest2010} and PSR J0348+0432
of $M = 2.01 \pm 0.04M_\odot$ \citep{Antoniadis:2013pzd})  larger than the typical observed pulsars.\footnote{Recall that General Relativity (GR) gives an upper limit  to NSs mass of $3.2 M_\odot$\Mov{ \cite{Rhoades:1974fn} using an extreme causal equation of state, although in Ref.~\cite{Brecher:1976zz} a more conservative limit of $4.7M_\odot$ is found. Gravitational waves (GW) observations accompanied by a short gamma-ray burst, assumed to be originated from a black hole center engine, lead to an upper mass limit of $\lesssim 2.2 M_\odot$ for NSs \cite{Margalit:2017dij,Rezzolla2017}.
However, should the short gamma-ray burst be originated instead from a magnetar center engine, the $2.2 M_\odot$ mass could become a lower limit. Note that the present heaviest pulsar PSR J0740+6620 reaches $2.14 M_\odot$ \cite{Stovall:2014gua,Lynch:2018zxo}. }
The theoretical lower limit to NSs mass is 0.1 $M_{\odot}$ but lepton-rich proto neutron stars are unbound below about 1 $M_{\odot}$ \cite{Lattimer2004}.}

Non-annihilating DM, among other results \citep{Li:2012ii,Sandin:2008db,Leung:2011zz,Xiang:2013xwa,Goldman:2013qla,Tolos2015}, yields the counterintuitive property of getting smaller
and less massive NSs, the more DM they accrete \citep{Sandin2009,Tolos2015}.
Studies have been performed on NSs admixed with mirror DM \citep{Sandin2009}, degenerate DM \citep{Leung2011}, and ADM \citep{Li2012}, finding that increasing the ratio of DM to normal matter yields NSs 
with smaller radii and masses. 
In Ref. \citep{Tolos2015} NSs and  White Dwarfs (WDs) matter admixed with 100 GeV ADM were studied, finding that planets-like objects could form. 
Those results were extended in \citep{Deliyergiyev2019} to particle masses in the range 1-500 GeV.

In this paper, 
based on the results obtained in our previous work \citep{Deliyergiyev2019}, we propose a testable galactic probe for DM existence in the form of the evolution of the pulsar mass towards the galactic centre (GC).
According to 
the discussion
above, NSs in increasingly DM rich environments should accrete more DM and thus
display a characteristic mass decrease, the closer they are to the galactic
centre. We use NSs because     
\begin{enumerate*}[label=(\arabic*)]
                            \item the very large baryon density inside NSs makes 
an interaction between baryons and DM following DM capture most likely;
                            \item the NSs strong gravitational force makes DM particles escape very unlikely, after they interact and loose energy.
                           \end{enumerate*}
This mass evolution is easier to test than other probes such as NS temperature time evolution with DM accretion, as discussed previously.

In Sec.~\ref{sec:Implementation} we discuss our model for DM
accretion in NS. In Sec.~\ref{sec:NSmassMW} we discuss, using simulations and observations, how to get a realistic dark matter halo, while  in Sec.~\ref{sec:NSmassChange} we compare the predicted change of NS 
mass,  moving towards the GC, with
observations. In Sec.~\ref{sec:GCmass} we discuss the use of pulsars in the GC, before to 
present our conclusions in Sec.~\ref{sec:Conclusions}.

\section{Accumulation of Dark Matter in Compact Objects}
\label{sec:Implementation} 

The equilibrium of ADM admixed onto a NS was studied in several papers, including \citep{Tolos2015,Deliyergiyev2019}, by means of the TOV equation 
for visible (or ordinary) matter (OM) and ADM, minimally coupled to gravity \citep{Deliyergiyev2019}. Among other results, it was shown that the compact object (CO) mass decreases as the accumulated DM increases.
Solving TOV equation 
only tells how much DM (as well as OM) 
a CO can contain, which is $\simeq 10^{-5} M_{\odot}$ for a particle mass of 200 GeV \citep{Deliyergiyev2019}. 
However, the process to provide that amount of DM to a given CO remains to be determined.
The acquisition of DM onto a NS  can be divided into three phases, i.e. 
\begin{enumerate*}[label=(\arabic*)] 
 \item the collapse of the protostar; 
 \item the star evolution until supernova explosion;
 \item the NS phase.
\end{enumerate*}

To our knowledge, there is only one simulation that estimates the DM accretion onto a NS \citep{Yang2011}, to construct a model that builds 
up the DM external to the NS.
Analytical studies provided an estimate of the accreted DM by NSs, making some simplifications, such as neglecting the  DM capture during the pre-NS phase \citep{Kouvaris2008,Kouvaris2013}, or considering progenitor phase capture comparable to the NS phase \citep{Kouvaris2010}. Focusing on the NS phase, several authors decomposed this phase in further stages \citep{Kouvaris2008,Kouvaris2010,deLavallaz2010,Yang2011,Kouvaris2011,Guver2014,Zheng2016}:
\begin{enumerate*}[label=(\arabic*)]
 \item DM capture in NS coming from DM-nucleon scattering;
 \item DM orbit decrease due to DM-neutron scattering\label{enu:DM-n};
 \item DM-DM interaction in NS.
\end{enumerate*}

The phase \ref{enu:DM-n} 
can lead to a Bose-Einstein condensate or a black hole (BH) formation \citep{Guver2014,Kouvaris2013}. Disregarding this phase, we can obtain the DM particle number evolution by
\begin{equation} \label{capt}
\frac{dN_{\rm dm}}{dt}=C_{\rm c}+C_{\rm s} N_{\rm dm} \ ,
\end{equation}
as determined in Eq.~(3.8) of Ref.~\citep{Guver2014}, where $C_{\rm c}$ is the DM-nucleon capture rate, and $C_{\rm s}$ is the DM self-interaction capture rate (see \citep{Guver2014}).
%
$C_{\rm c}$ is given in Ref.~\citep{Kouvaris2011}, assuming a Maxwellian DM distribution, and taking into account general relativistic corrections:
\citep[see][]{Kouvaris2008,Kouvaris2010,Kouvaris2011,Guver2014}.
%
%

As shown by Ref.~\cite{Kouvaris2013}, in the framework of a spherically symmetric accretion scenario for a typical NS of mass $1.4 M_\odot$ and R=10 km, the total accreted mass 
is
\begin{equation} 
M_{\rm acc}= 1.3 \times 10^{43} \left( \frac{\rho_{\rm dm}}{\rm 0.3 \, GeV/cm^3} \right) \left(\frac{\rm t}
{\rm Gyr} \right) f \,\,\, \rm GeV,
\label{eq:Kouv2013}
\end{equation}
where $\rho_{\rm dm}$ is the local DM density, 
t is the accretion time of DM by the NS and $f$ gives the NS particle fraction undergoing scatterings while in the NS, that is set 
$f=1$ for scattering cross section $\sigma_{\rm dm} >10^{-45}$ $\rm cm^2$, or $f=0.45 \, \sigma_{\rm dm}/\sigma_{\rm crit}$, with $\sigma_{\rm crit} \simeq 6 \times 10^{-46}$ $\rm cm^2$, for a homogeneous NS \citep[see][for a detailed calculation]{Kouvaris2008}\footnote{Note that in usually inhomogenehous NS, $f$ will be larger than the proposed 
estimate \citep{Kouvaris2008}}.
 From  XENON1T \citep{Aprile2017}, $\sigma_{\rm dm}$  for a particle mass of some hundreds of GeV is $\simeq 10^{-45} cm^2$, leading to $f \simeq 1$. Note that Eq.~(\ref{eq:Kouv2013}) is an underestimation of a factor $\simeq 10$, since the accretion during the NS progenitor phase, of the same order as in the NS phase \citep{Kouvaris2010} (factor of 2), and the accretion coming from DM self-interaction \citep{Guver2014} are not taken into account, and finally because in this paper, we will use NSs having mass $2 M_\odot$\footnote{Then 
 Eq.~(\ref{eq:Kouv2013}) must also multiplied by a factor 2.05.}.
%
%
The time $t$ changes from pulsar to pulsar. Pulsars can be very young (e.g., CRAB pulsar), very old, $>10^{10}$ yrs (PSR J1518+4904, PSR J1829+2456) \citep{Wong2010}, $\simeq 2\rm Gyr$ (PSR J1811-1736) \citep{Wong2010}, or present intermediate ages, e.g. some $10^8$ yrs (PSR B1534+12
, PSR J0737-3039, PSR J1756-2251, etc) \citep{Lorimer1998,Wong2010}. In our calculation, we used 
$t=10 \rm Gyr$\footnote{ In what follows we discuss 
the effect of a change in total accretion time t.}. We chose this value because the number of old
neutron stars ($10^9$, $10^{10}$ yrs) increases going
towards the galactic center \cite{Taani:2012xp,Ofek:2009wt,Wei:2010}  with a maximum around 3 kpc (note that the
majority of pulsars are located at $r>3$ kpc -- ATNF catalogue). It is highly probable that the trend continues going towards the galactic center. 

Moreover,  NSs and Supernovae numbers increase going towards the galactic center \cite{Sartore:2009wn,Dragicevich:1999}, inducing a larger
probability to find older objects close to the GC. These are other obvious incentives to eagerly wait for 
new observations finding objects closer to the GC.

We obtain DM accretion $\simeq 10^{-11} {\rm M}_{\odot}$  for a typical NS in the solar neighbourhood
, which is in agreement with the capture rates of Refs~\citep{Kouvaris2013,Zhong2012,Zheng2016,Guver2014} and the  results from  \citep{Kouvaris2011}, but below the estimates from the DM accumulated using TOV
. A better agreement between the accreted DM mass and the accumulated DM  mass coming from TOV 
is obtained for NSs located in Superdense DM clumps, Ultra Compact mini-haloes \citep{Berezinsky2013}, and close to the GC.
%
%
We recall that another approach to DM accretion is proposed by 
\cite{Chang2018}.

%
%

\section{Dark Matter in the Milky Way}
\label{sec:NSmassMW}

The nature of the the Milky Way (MW) density profile is not known from observations \citep{Ullio2010,Weber2010}, with some simulations predicting cuspy profiles \citep{Navarro1997}, while some more recent ones finding a 
flattening towards the GC \citep{Stadel2009,Navarro2010}. Consequently, the determination of the DM environment in  which NSs are embedded in MW, and the estimation of DM accreted by them, poses problem. In this paper, we use the Einasto profile, as it yields a good description of recent simulations  \citep{Stadel2009}. Note that such profile, obtained in DM-only simulations, does not consider baryon effects. Moreover, as COs accrete DM according to their 
environment, locally dense environments lead to larger DM content, which gives rise to smaller mass COs.

\begin{figure*}[!hpt]
\hspace{-2.5cm}\includegraphics[scale=0.41]{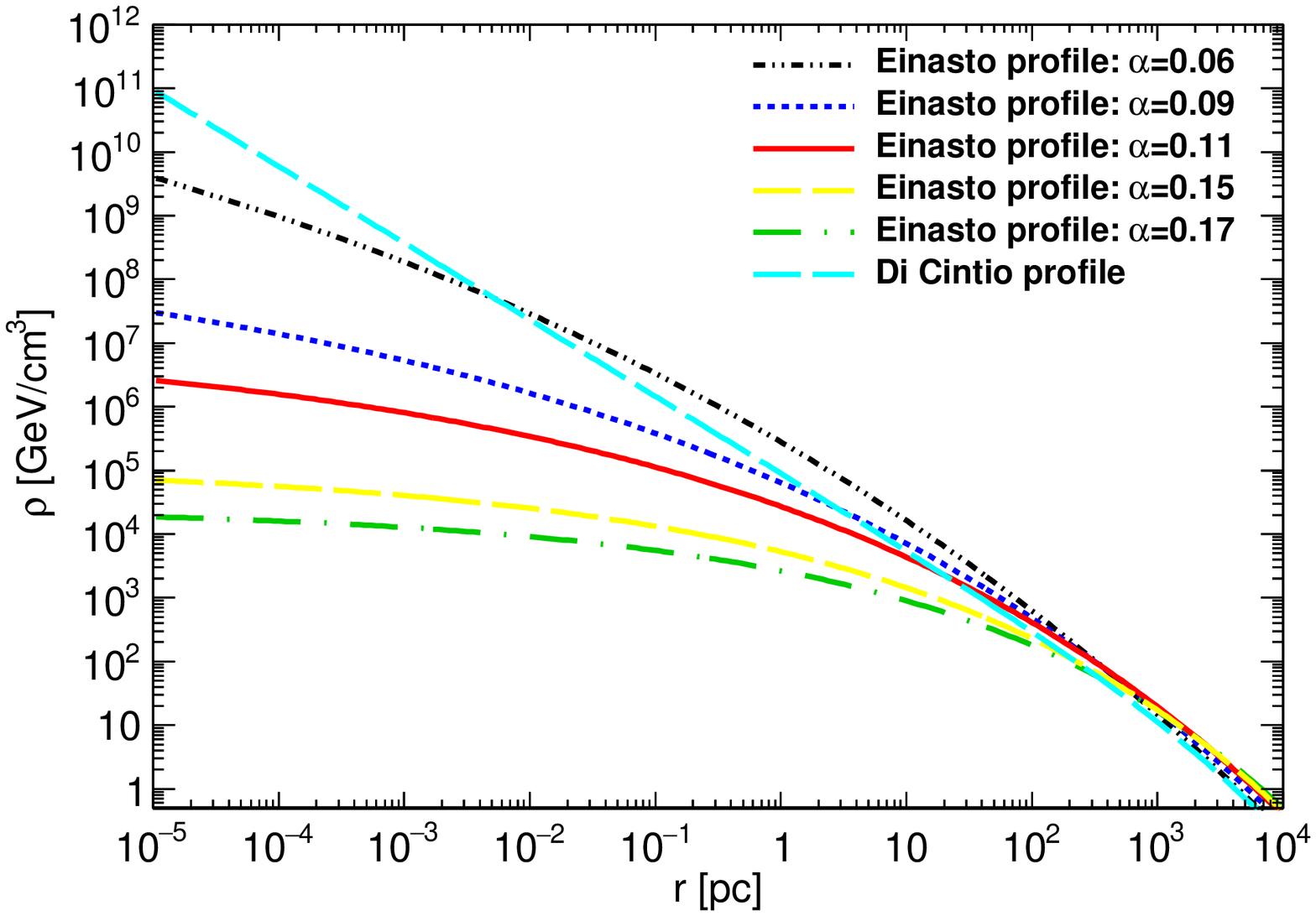}
\includegraphics[scale=0.41]{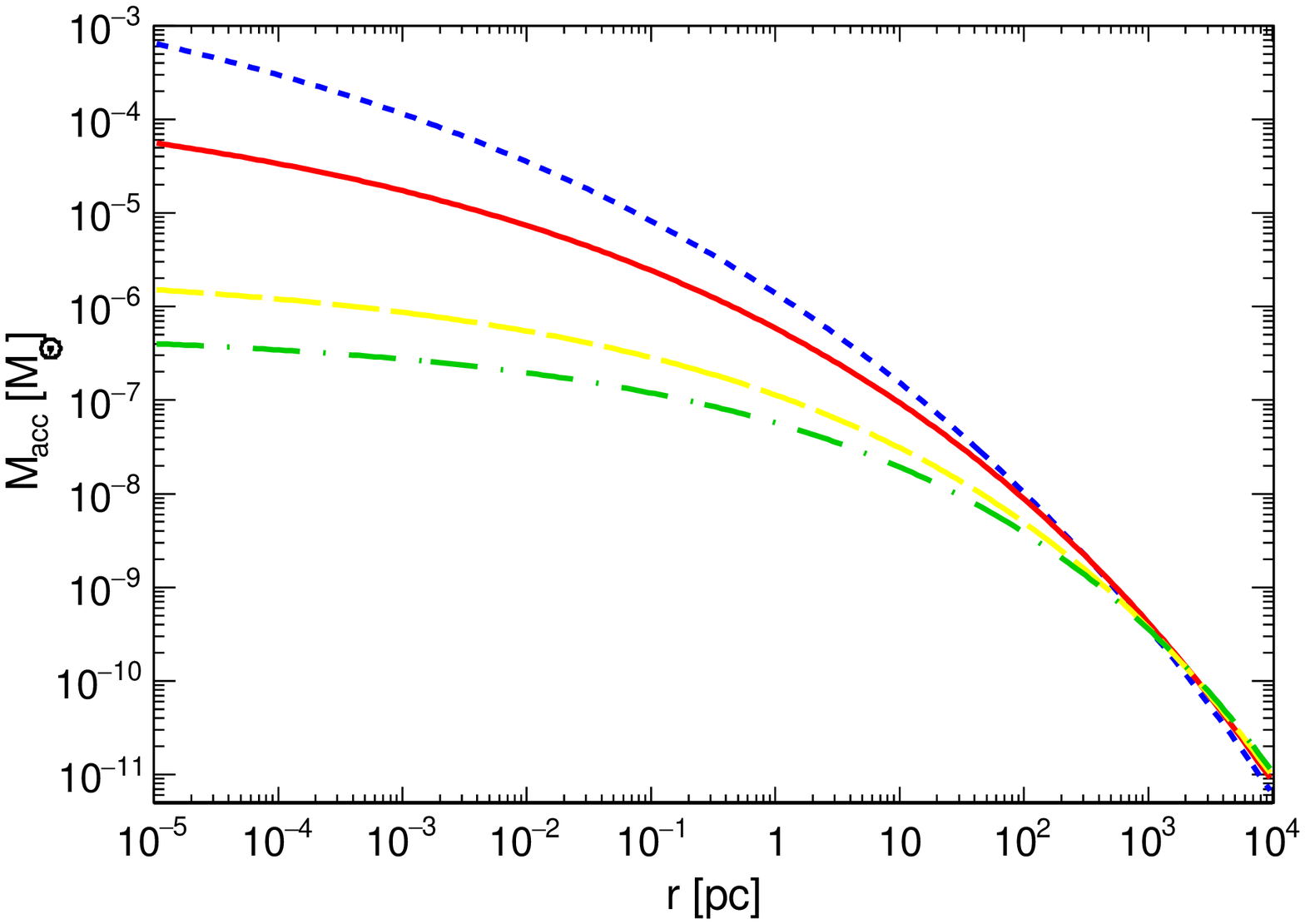}
 \caption{(Left panel) Einasto profile for $\alpha=0.06,0.09,0.11,0.15,0.17$ (black dot dot dashed, blue dotted, red solid, yellow long dashed, and green dot dashed lines, respectively). The cyan dashed line is the Di Cintio profile \cite{DiCintio2014}  (labelled DC14 in the text). (Right panel) The accreted mass according to 
Kouvaris formula (Eq.~(\ref{eq:Kouv2013})). In this plot we do not consider the accreted mass corresponding to the Di Cinto profile, nor to 
the Einasto profile with $\alpha=0.06$, 
in order to be conservative in our calculations. 
}
\label{fig:einasto}
\end{figure*}

In the left panel of Fig.~\ref{fig:einasto}, the Einasto DM density profile is plotted for different values of the parameters (see the figure caption for details),
\begin{equation}
\rho=\rho_{{-2}}{{\rm e}^{-2\,{\frac {1}{\alpha} \left[  \left( {\frac {r}{r_{
{-2}}}} \right) ^{\alpha}-1 \right] }}},\label{eq:einasto}
\end{equation}
where $\alpha$ gives the degree of cuspiness of the profile,  and $r_{-2}$ indicates the distance at fixed slope $\frac{d \ln{\rho}}{d \ln{r}}=-2$, where the density reaches $\rho_{{-2}}$. 
%
%

In order to fix the three free parameters of the Einasto profile, there are two possibilities:
\begin{enumerate*}[label=(\arabic*)] 
 \item simulations; 
 \item MW observations.
\end{enumerate*}
 Concerning simulations, the value of $\alpha$, lies in the range $0.12<\alpha<0.22$ \citep{Iocco2011}, while more recently Ref.~\citep{Udrescu2018,Dutton2014} found that $\alpha \simeq 0.15$ for a halo having the mass of the MW,
in agreement with other DM-only simulations (no baryons account) 
and MW observations. However, baryons affect the density profile by
\begin{enumerate*}[label=(\arabic*)]
 \item steepening it with adiabatic contraction \label{enu:adiabCont} \citep{Gnedin2004,Gustafsson2006,Pedrosa2009,Duffy2010};
 \item flattening it with supernovae feedback or similar effects \citep{DelPopolo2010,DiCintio2014,DelPopolo2016a,DelPopolo2017}, such as dynamical friction \citep{DelPopolo2009,DelPopolo2016a}.
\end{enumerate*}

Process \ref{enu:adiabCont} dominates MW-type galaxies and, as a consequence, DM-only simulations need a
correction as in Ref.~\cite{Prada2004}, that was later confirmed by hydro-dynamical simulations \cite{DiCintio2014}, which showed that the inner slope of the density profile is steepened as $\left.\frac{d\ln\rho}{d\ln r}\right|_{r\to0} \simeq -1.2$ \cite{Pedrosa2009}. This result translates into a reduction of the inner slope from 
$\alpha= 0.15$, given in DM-only simulations \citep{Udrescu2018,Dutton2014}, to $\alpha= 0.11$  \citep{Cirelli2010}, which 
we choose as our fiducial value. Due to adiabatic contraction,  
the concentration parameter changes from values $\simeq 8.5$ of DM-only simulations \cite{Maccio2008,Duffy2008} to a value two times larger, $18-20$ \citep{DiCintio2014}, in agreement with observations of the MW \cite{Ullio2010,Deason2012,Nesti2013}, producing a reduction of $r_{-2}$. For a halo having a virial mass $M_{\rm vir}$, and virial radius $r_{\rm vir}$, $r_{-2}=r_{\rm vir}/c$. Then, taking into account baryons effects, for MW-like mass halo, $r_{-2} \simeq 10$ kpc, in agreement with SPH simulations \citep{DiCintio2014}, and 2 times smaller than N-body only simulations (e.g. \citep{Udrescu2018}). In the following, we use as fiducial value 15 kpc. In order to get the normalization, i.e. the third parameter $\rho_{-2}$, we use Eq.~(\ref{eq:einasto}) with the previous given values of $\alpha$, and $r_{-2}$, and assuming that the local density, $\rho(R_{\odot})$, being $R_{\odot}$ the distance Sun-GC,  has the value estimated by \citep{Pato2015}, $\rho(R_{\odot})=0.420^{+0.019}_{-0.021}$, using their value of $R_{\odot}$. 
The Einasto model with those 
parameters, and all the quantities related to it, is 
plotted in red (solid line). In the paper, we will also consider a range of parameterization differing from the fiducial case.

As previously reported, observations, within the limits of their uncertainties, allow to constraint the Einasto profile. For a given $\alpha$, mass constraints, e.g.  $M(60\, \rm kpc)=(4 \pm 0.7) \times 10^{11} M_{\odot}$ \citep{Xue2008,Bernal2012}, together with, e.g., the \citep{Pato2015} local density $\rho_\odot$, allows to fix the density profile parameters \citep{Cirelli2010}.\footnote{($\simeq 10 \rm kpc$, $\rho_{-2} \simeq 0.3 \rm GeV/cm^3$, 0.11).} 
$\gamma$-rays observations \citep{Bernal2012} have also been used to constraint the Einasto profile parameters, together with 
mass modelling \citep{Catena2010,Karukes2019}
obtaining constraints in agreement with those of \citep{Iocco2011}, and with our fiducial case: $0.10<\alpha<0.22$, $8<r_{-2}<30$ kpc.

%
%
\begin{figure*}[!hpt]
 \hspace{-3.5cm}\includegraphics[scale=0.91]{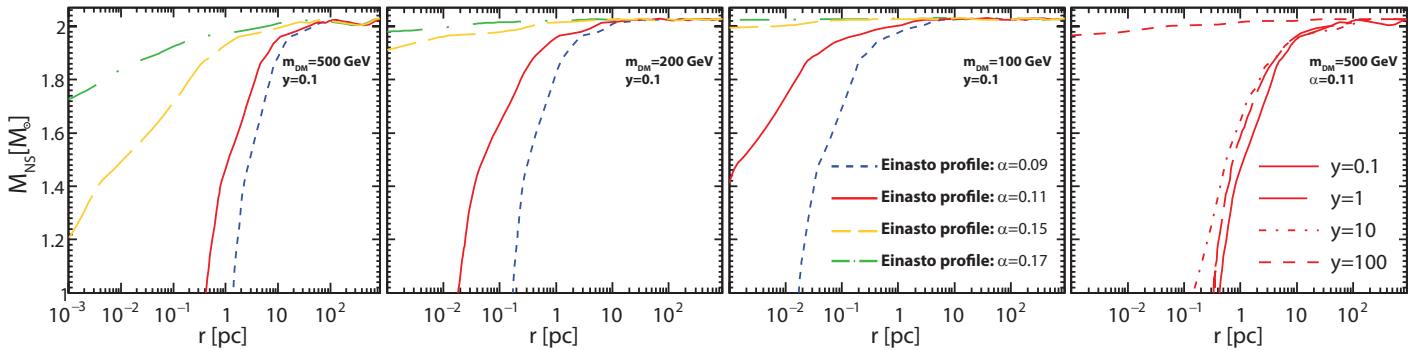}
\caption{
The changes of the NS mass from DM accumulation as a function of the NS distance to GC, 
for a particle mass $M=500$ (leftmost), 200 (center left) and 100 (center right) $\rm GeV$,  with $y=0.1$ and $\alpha=0.09,0.11,0.15,0.17$ (blue dotted, red solid, yellow long dashed
, and green dot dashed). The rightmost panel shows the role of the interaction strength for 
particle mass $M=500 \rm GeV$, $\alpha=0.11$, and $y=0.1, 1, 10, 100$ (solid, long dashed, dash dotted, and long dashed dotted lines). }
 \label{fig:M_NS_45}
\end{figure*}

The right panel of Fig.~\ref{fig:einasto} shows the corresponding DM accreted using the formula given in Eq.~(\ref{eq:Kouv2013}) from Ref.~\citep{Kouvaris2013}. The blue dotted, red solid, yellow long dashed, and green dot dashed lines correspond to $\alpha=0.09, 0.11, 0.15, 0.17$. This means, that while a NS located at the Sun neighborhood will accrete $\simeq 10^{-11} {\rm M}_{\odot}$, one located at $10^{-5} (0.1)$ pc will accrete $6.5\times10^{-4} (3.4\times 10^{-6}) {\rm M}_{\odot}$, in the case $\alpha=0.09$.
%
%
\section{Mass Change of Neutron Stars}\label{sec:NSmassChange}
Once the accreted DM mass, $M_{\rm acc}$, as a function of the distance to the GC is determined from the right panel 
of Fig.~\ref{fig:einasto}, we can determine the corresponding mass change of the NS
. In order to do so, we use Fig.~10 
of our previous work \citep{Deliyergiyev2019}, where the maximum mass of NSs was obtained as a function of the DM mass inside the NS, $M_{\rm DM}$, for the DM weakly interacting case, $y=0.1$\footnote{ The interaction strength is expressed in terms of the ratio of the DM fermion mass $m_f$, and scale of interaction $m_I$, $y=m_f/m_I$. This can be converted to usual units: one can estimate the cross section of DM self-interaction, taking the mass of the DM particle $m_f$ in units of GeV, as \begin{align}
        \sigma =& \frac{1}{4\pi}\frac{m_f^2}{m_I^4}
               = \frac{y^4}{m_f^2} 3.2\times 10^{-27} \textrm{cm}^2  &
               \rightarrow \sigma/m_f =& \frac{y^4}{m_f^3} 1.8\times 10^{-3} \textrm{cm}^2/g\label{eq:yCrossSection}
               \end{align}
}. Moreover, we obtained similar plots for larger values of the interaction parameter $y$, $y=1, 10, 100$.
Since $M_{\rm DM}$ must be equal to the accreted mass $M_{\rm acc}$, we find a relation between the total mass of the NS, $M_{\rm NS}$ ($M_{\rm T}$ in the notation of \citep{Deliyergiyev2019})  and the distance from the GC.
The result is plotted in Fig.~\ref{fig:M_NS_45}. Its leftmost panel displays the change in mass of a NS moving towards the MW halo center (colour and line coding as in Fig.~\ref{fig:einasto}), for a particle mass, $M=500 \,\rm GeV$, and $y=0.1$.  Our reference model ($\alpha=0.11$, red solid line) shows a NS mass change from 2 to 1 $M_{\odot}$ at 0.4 pc\footnote{Note that reducing the accumulation time, t, shifts the curves 
inward, i.e. towards smaller radii, by a same 
factor
.}, while in the case $\alpha=0.09$ (blue dotted line) the same change is observed at 1.35 pc. The other two cases show a slower mass change. The yellow long dashed line ($\alpha=0.15$), and the green dot dashed line ($\alpha=0.17$) show that the mass reduces from 2 to 1.2 $M_{\odot}$, and from 2 to 1.7 $M_{\odot}$, at $10^{-3} $ pc, respectively.
%
%
%
The next two panels show how the mass changes with decreasing the DM particle mass ($M=200$ GeV, centre left, $100$ GeV, centre right). Finally the rightmost plot shows the effect of the interaction strength for our reference profile ($\alpha=0.11$), and  from right to left, in the cases $y=0.1, 1, 10, 100$. 
Strongest interaction ($y=100$) produces very small mass changes,
while the weaker the interaction,  the larger the change is.

All the above results are obtained with the conservative assumption
of using a density profile from DM-only simulations, shallower than in recent hydrodynamic simulations \cite{DiCintio2014}.

As the right panel of Fig.~\ref{fig:einasto}\;\;shows, we excluded from the analysis the $\alpha=0.06$ Einasto profile, and even the DC14 profile \cite{DiCintio2014}, a realistic profile calculated with hydro-dynamical simulations, to remain conservative. 
In the same line, a recent paper appearing after our results were publicised \cite{Ivanytskyi:2019wxd}, using the steeper NFW profile, that entails a central density 600 times higher than our conservative profile, is reporting a DM to total mass fraction of order 1\% for NSs at a distance around the centre of the order of kpc, while our model predicts at 1kpc a fraction of $7.4\times 10^{-4}\%$, neglecting possible spikes due to the BH.
For precision's sake, most probably, the best description of 
the density profile close to the galactic center is DC14 \cite[and references therein]{DiCintio2014}, since it describes very well, and shows very good agreement with, the profile of observed galaxies. This would induce the central density of a factor $3\times10^4$ larger than what we have plotted. 
Such conjecture is reinforced, as several papers are agreeing on the fact that the density close to the galactic center is of the order of, or even much larger ($10^4$ times larger) than, that indicated by DC14 \cite[i.e.][]{BertoneMerritt2005,deLavallaz2010,Gondolo1999}. Furthermore, Ref~\cite{Sandick2018} expect DM density spikes in the GC, whereas Lacroix \cite{Lacroix2018} does not exclude a spike with radius smaller than a few tens of parsecs for cuspy outer halos. Recently, \cite{Bennewitz2019} examined the spikes and gave several references discussing their existence \cite[e.g.][]{Fields2014,ShapiroShelton2016}.

Although 
we made the choice of a more conservative density profile, we note that the orbital dynamics of PSR B1257+12  \cite{Iorio2010}, together with the accretion predictions \cite{BertoneMerritt2005} allows much larger DM accumulation in NS than in our present work: up to 10\% of a NS mass, in agreement with DC14 \citep{DiCintio:2014xia}. 
%
%
Moreover, complex astrophysical phenomena are occurring on sub-parsec scale near the GC, such as 
DM particles gravitational scattering by stars and capture in the supermassive BH, together with highly enhanced central density from the supermassive BH formation \cite{Gondolo1999,BertoneMerritt2005}. Thus, more accurate density profiles can be introduced \citep{BertoneMerritt2005,deLavallaz2010}.

\section{Mass Determination of Neutron Stars close to the Galactic Center}
\label{sec:GCmass}

At this stage, the remaining 
question is whether the mass change of NSs can be determined by observations.
Despite the hundreds, or up to thousands, 
pulsars theoretically expected in the GC \cite{Pfahl2003,Wharton2011,Chennamangalam2013}, only six have been detected in the Galaxy inner $30'$ \cite{Eatough2013,Mori2013,Kennea2013,Shannon2013}. In particular, the J1745-2900  \cite{Eatough2013,Mori2013,Kennea2013,Shannon2013}, a transient magnetar, is located 0.1 pc from the GC.
This "missing pulsars" problem can be understood by hyper-strong interstellar scattering \cite{Lazio1998}, more complex scattering models, stellar population synthesis arguments, or pulsar emission suppression mechanisms \cite{Bower2018}. 


Estimates for SKA (Square Kilometer Array) \cite{FaucherGiguere2010} advocate detection of a $L_{1000} \simeq 0.7\, \rm mJy \, kpc^2$ 5-ms MSP (millisecond pulsar) at the GC with a ratio signal/noise of $S/N = 10$ and spectral index $\alpha=-1$. A few pulsar-black hole binaries are also estimated \cite{FaucherGiguere2010} in the inner parsec, while \cite{Chennamangalam2013} propose a conservative upper limit $\simeq 200$, albeit 
\citep{Rajwade2017} predicts that up to 52 canonical pulsars could be observed, and  
10000 MSPs, by SKA and Next Generation Very Large Array (ngVLA) surveys \citep{Murphy:2018vxa,Keane2015}. 
In addition
, the factor of 10 improvement in sensitivity at high frequencies promised by the ngVLA \citep{Murphy:2018vxa} will be an unprecedented probe of General Relativity and black hole physics, making dramatic improvements on the detection of pulsars close to the GC. These promisses are already in motion with the Event Horizon Telescope (EHT)\citep{Luminet2019} whose team recently released the first image of the M87 black hole.

After a pulsar is detected, the mass needs to be determined. Mass measurements can be obtained in different ways \cite{Engineer1998, Watts:2016uzu}, and recently, an interesting new technique was added to the several already known
, based on pulsar glitch data to constrain superfluid and nuclear EoS models \cite{Ho:2015vza}. The upcoming SKA \citep{Konar:2016lgc}, Athena \citep{Barcons:2012zb,Athena:2014cdf}, NICER \citep{NICER:2012} and eXTP \citep{Watts:2018iom} observatories are expected to offer precise measurements of the masses, together with radii, pinning down the composition of NS.
%
%

\section{Conclusions}

\label{sec:Conclusions}

We have shown that the NS mass should reflect the changes of the DM environment in the MW (Fig.~\ref{fig:M_NS_45}). This is done taking into account the DM accretion of NSs \citep[see][for details]{Deliyergiyev2019}, as it changes because of the increase of DM content \citep{Kouvaris2013} when we move towards the GC \citep{Bernal2012}. 
This allows us to propose that the evolution of the pulsar masses towards the GC of the MW can be a probe of the existence of DM. In fact, the decrease of the NS mass for NSs located closer and closer 
to the GC would put constraints on the characteristics of the Galaxy halo dark matter profile, on the dark matter particle mass, and on the self-interaction strength. Such changes are expected to be observed in the near future in telescopes such as ngVLA, SKA, Athena, NICER or eXTP \cite{Murphy:2018vxa,Konar:2016lgc,Barcons:2012zb,Athena:2014cdf,NICER:2012,Watts:2018iom}.



\section*{Acknowledgments}
M.D. work was supported by the Chinese Academy of Sciences Presidents International Fellowship Initiative Under Grant No. 2016PM043 and by the Polish National Science Centre (NCN) grant 2016/23/B/ST2/00692. MLeD acknowledges the financial support by Lanzhou University starting fund and the Fundamental Research Funds for the
Central Universities (Grant No.lzujbky-2019-25).
L.T. acknowledges support from the FPA2016-81114-P Grant from Ministerio de Ciencia, Innovacion y Universidades, Heisenberg Programme of the Deutsche Forschungsgemeinschaft under the Project Nr. 383452331 and PHAROS COST Action CA16214.

\bibliographystyle{elsarticle-num}

\bibliography{biblioNS,old_MasterBib2}

\begin{thebibliography}{100}
\expandafter\ifx\csname url\endcsname\relax
  \def\url#1{\texttt{#1}}\fi
\expandafter\ifx\csname urlprefix\endcsname\relax\def\urlprefix{URL }\fi
\expandafter\ifx\csname href\endcsname\relax
  \def\href#1#2{#2} \def\path#1{#1}\fi

\bibitem{Betoule:2014frx}
M.~Betoule, et~al., {Improved cosmological constraints from a joint analysis of
  the SDSS-II and SNLS supernova samples}, Astron. Astrophys. 568 (2014) A22.
\newblock \href {http://arxiv.org/abs/1401.4064} {\path{arXiv:1401.4064}},
  \href {http://dx.doi.org/10.1051/0004-6361/201423413}
  {\path{doi:10.1051/0004-6361/201423413}}.

\bibitem{Ade:2013zuv}
P.~A.~R. Ade, et~al., {Planck 2013 results. XVI. Cosmological parameters},
  Astron. Astrophys. 571 (2014) A16.
\newblock \href {http://arxiv.org/abs/1303.5076} {\path{arXiv:1303.5076}},
  \href {http://dx.doi.org/10.1051/0004-6361/201321591}
  {\path{doi:10.1051/0004-6361/201321591}}.

\bibitem{Chatrchyan:2012me}
S.~Chatrchyan, et~al., {Search for dark matter and large extra dimensions in
  monojet events in $pp$ collisions at $\sqrt{s}=7$ TeV}, JHEP 09 (2012) 094.
\newblock \href {http://arxiv.org/abs/1206.5663} {\path{arXiv:1206.5663}},
  \href {http://dx.doi.org/10.1007/JHEP09(2012)094}
  {\path{doi:10.1007/JHEP09(2012)094}}.

\bibitem{ATLAS:2012ky}
G.~Aad, et~al., {Search for dark matter candidates and large extra dimensions
  in events with a jet and missing transverse momentum with the ATLAS
  detector}, JHEP 04 (2013) 075.
\newblock \href {http://arxiv.org/abs/1210.4491} {\path{arXiv:1210.4491}},
  \href {http://dx.doi.org/10.1007/JHEP04(2013)075}
  {\path{doi:10.1007/JHEP04(2013)075}}.

\bibitem{Agnese:2014aze}
R.~Agnese, et~al., {Search for Low-Mass Weakly Interacting Massive Particles
  with SuperCDMS}, Phys. Rev. Lett. 112~(24) (2014) 241302.
\newblock \href {http://arxiv.org/abs/1402.7137} {\path{arXiv:1402.7137}},
  \href {http://dx.doi.org/10.1103/PhysRevLett.112.241302}
  {\path{doi:10.1103/PhysRevLett.112.241302}}.

\bibitem{Angloher:2011uu}
G.~Angloher, et~al., {Results from 730 kg days of the CRESST-II Dark Matter
  Search}, Eur. Phys. J. C72 (2012) 1971.
\newblock \href {http://arxiv.org/abs/1109.0702} {\path{arXiv:1109.0702}},
  \href {http://dx.doi.org/10.1140/epjc/s10052-012-1971-8}
  {\path{doi:10.1140/epjc/s10052-012-1971-8}}.

\bibitem{Felizardo:2011uw}
M.~Felizardo, et~al., {Final Analysis and Results of the Phase II SIMPLE Dark
  Matter Search}, Phys. Rev. Lett. 108 (2012) 201302.
\newblock \href {http://arxiv.org/abs/1106.3014} {\path{arXiv:1106.3014}},
  \href {http://dx.doi.org/10.1103/PhysRevLett.108.201302}
  {\path{doi:10.1103/PhysRevLett.108.201302}}.

\bibitem{Klasen:2015uma}
M.~Klasen, M.~Pohl, G.~Sigl, {Indirect and direct search for dark matter},
  Prog. Part. Nucl. Phys. 85 (2015) 1--32.
\newblock \href {http://arxiv.org/abs/1507.03800} {\path{arXiv:1507.03800}},
  \href {http://dx.doi.org/10.1016/j.ppnp.2015.07.001}
  {\path{doi:10.1016/j.ppnp.2015.07.001}}.

\bibitem{Akerib:2013tjd}
D.~S. Akerib, et~al., {First results from the LUX dark matter experiment at the
  Sanford Underground Research Facility}, Phys. Rev. Lett. 112 (2014) 091303.
\newblock \href {http://arxiv.org/abs/1310.8214} {\path{arXiv:1310.8214}},
  \href {http://dx.doi.org/10.1103/PhysRevLett.112.091303}
  {\path{doi:10.1103/PhysRevLett.112.091303}}.

\bibitem{Ahmed:2010wy}
Z.~Ahmed, et~al., {Results from a Low-Energy Analysis of the CDMS II Germanium
  Data}, Phys. Rev. Lett. 106 (2011) 131302.
\newblock \href {http://arxiv.org/abs/1011.2482} {\path{arXiv:1011.2482}},
  \href {http://dx.doi.org/10.1103/PhysRevLett.106.131302}
  {\path{doi:10.1103/PhysRevLett.106.131302}}.

\bibitem{Bernabei:2010mq}
R.~Bernabei, et~al., {New results from DAMA/LIBRA}, Eur. Phys. J. C67 (2010)
  39--49.
\newblock \href {http://arxiv.org/abs/1002.1028} {\path{arXiv:1002.1028}},
  \href {http://dx.doi.org/10.1140/epjc/s10052-010-1303-9}
  {\path{doi:10.1140/epjc/s10052-010-1303-9}}.

\bibitem{Aalseth:2010vx}
C.~E. Aalseth, et~al., {Results from a Search for Light-Mass Dark Matter with a
  P-type Point Contact Germanium Detector}, Phys. Rev. Lett. 106 (2011) 131301.
\newblock \href {http://arxiv.org/abs/1002.4703} {\path{arXiv:1002.4703}},
  \href {http://dx.doi.org/10.1103/PhysRevLett.106.131301}
  {\path{doi:10.1103/PhysRevLett.106.131301}}.

\bibitem{Aprile:2012nq}
E.~Aprile, et~al., {Dark Matter Results from 225 Live Days of XENON100 Data},
  Phys. Rev. Lett. 109 (2012) 181301.
\newblock \href {http://arxiv.org/abs/1207.5988} {\path{arXiv:1207.5988}},
  \href {http://dx.doi.org/10.1103/PhysRevLett.109.181301}
  {\path{doi:10.1103/PhysRevLett.109.181301}}.

\bibitem{Conrad:2014tla}
J.~Conrad,
  \href{http://inspirehep.net/record/1326617/files/arXiv:1411.1925.pdf}{{Indirect
  Detection of WIMP Dark Matter: a compact review}}, in: {Interplay between
  Particle and Astroparticle physics (IPA2014) London, United Kingdom, August
  18-22, 2014}, 2014.
\newblock \href {http://arxiv.org/abs/1411.1925} {\path{arXiv:1411.1925}}.
\newline\urlprefix\url{http://inspirehep.net/record/1326617/files/arXiv:1411.1925.pdf}

\bibitem{Dai:2009ik}
D.-C. Dai, D.~Stojkovic, {Neutralino dark matter stars can not exist}, JHEP 08
  (2009) 052.
\newblock \href {http://arxiv.org/abs/0902.3662} {\path{arXiv:0902.3662}},
  \href {http://dx.doi.org/10.1088/1126-6708/2009/08/052}
  {\path{doi:10.1088/1126-6708/2009/08/052}}.

\bibitem{Kouvaris:2015rea}
C.~Kouvaris, N.~G. Nielsen, {Asymmetric Dark Matter Stars}, Phys. Rev. D92~(6)
  (2015) 063526.
\newblock \href {http://arxiv.org/abs/1507.00959} {\path{arXiv:1507.00959}},
  \href {http://dx.doi.org/10.1103/PhysRevD.92.063526}
  {\path{doi:10.1103/PhysRevD.92.063526}}.

\bibitem{Bertolami:2007zm}
O.~Bertolami, F.~Gil~Pedro, M.~Le~Delliou, {Dark Energy-Dark Matter Interaction
  and the Violation of the Equivalence Principle from the Abell Cluster A586},
  Phys. Lett. B654 (2007) 165--169.
\newblock \href {http://arxiv.org/abs/astro-ph/0703462}
  {\path{arXiv:astro-ph/0703462}}, \href
  {http://dx.doi.org/10.1016/j.physletb.2007.08.046}
  {\path{doi:10.1016/j.physletb.2007.08.046}}.

\bibitem{LeDelliou:2007am}
M.~Le~Delliou, O.~Bertolami, F.~Gil~Pedro, {Dark Energy-Dark Matter Interaction
  from the Abell Cluster A586 and violation of the Equivalence Principle}, AIP
  Conf. Proc. 957 (2007) 421--424.
\newblock \href {http://arxiv.org/abs/0709.2505} {\path{arXiv:0709.2505}},
  \href {http://dx.doi.org/10.1063/1.2823818} {\path{doi:10.1063/1.2823818}}.

\bibitem{Bertolami:2008rz}
O.~Bertolami, F.~Gil~Pedro, M.~Le~Delliou, {Dark Energy-Dark Matter Interaction
  from the Abell Cluster A586}, EAS Publ. Ser. 30 (2008) 161--167.
\newblock \href {http://arxiv.org/abs/0801.0201} {\path{arXiv:0801.0201}},
  \href {http://dx.doi.org/10.1051/eas:0830019}
  {\path{doi:10.1051/eas:0830019}}.

\bibitem{Bertolami:2007tq}
O.~Bertolami, F.~G. Pedro, M.~Le~Delliou, {The Abell Cluster A586 and the
  Equivalence Principle}, Gen. Rel. Grav. 41 (2009) 2839--2846.
\newblock \href {http://arxiv.org/abs/0705.3118} {\path{arXiv:0705.3118}},
  \href {http://dx.doi.org/10.1007/s10714-009-0810-1}
  {\path{doi:10.1007/s10714-009-0810-1}}.

\bibitem{Bertolami:2012yp}
O.~Bertolami, F.~Gil~Pedro, M.~Le~Delliou, {Testing the interaction of dark
  energy to dark matter through the analysis of virial relaxation of clusters
  Abell Clusters A586 and A1689 using realistic density profiles}, Gen. Rel.
  Grav. 44 (2012) 1073--1088.
\newblock \href {http://arxiv.org/abs/1105.3033} {\path{arXiv:1105.3033}},
  \href {http://dx.doi.org/10.1007/s10714-012-1327-6}
  {\path{doi:10.1007/s10714-012-1327-6}}.

\bibitem{Abdalla:2007rd}
E.~Abdalla, L.~R.~W. Abramo, L.~Sodre, Jr., B.~Wang, {Signature of the
  interaction between dark energy and dark matter in galaxy clusters}, Phys.
  Lett. B673 (2009) 107--110.
\newblock \href {http://arxiv.org/abs/0710.1198} {\path{arXiv:0710.1198}},
  \href {http://dx.doi.org/10.1016/j.physletb.2009.02.008}
  {\path{doi:10.1016/j.physletb.2009.02.008}}.

\bibitem{Abdalla:2009mt}
E.~Abdalla, L.~R. Abramo, J.~C.~C. de~Souza, {Signature of the interaction
  between dark energy and dark matter in observations}, Phys. Rev. D82 (2010)
  023508.
\newblock \href {http://arxiv.org/abs/0910.5236} {\path{arXiv:0910.5236}},
  \href {http://dx.doi.org/10.1103/PhysRevD.82.023508}
  {\path{doi:10.1103/PhysRevD.82.023508}}.

\bibitem{Delliou:2014awa}
M.~Le~Delliou, R.~J.~F. Marcondes, G.~B. Lima~Neto, E.~Abdalla, {Non-virialized
  clusters for detection of dark energy–dark matter interaction}, Mon. Not.
  Roy. Astron. Soc. 453~(1) (2015) 2--13.
\newblock \href {http://arxiv.org/abs/1411.5863} {\path{arXiv:1411.5863}},
  \href {http://dx.doi.org/10.1093/mnras/stv1561}
  {\path{doi:10.1093/mnras/stv1561}}.

\bibitem{Tolman:1939jz}
R.~C. Tolman, {Static solutions of Einstein's field equations for spheres of
  fluid}, Phys. Rev. 55 (1939) 364--373.
\newblock \href {http://dx.doi.org/10.1103/PhysRev.55.364}
  {\path{doi:10.1103/PhysRev.55.364}}.

\bibitem{Oppenheimer:1939ne}
J.~R. Oppenheimer, G.~M. Volkoff, {On Massive neutron cores}, Phys. Rev. 55
  (1939) 374--381.
\newblock \href {http://dx.doi.org/10.1103/PhysRev.55.374}
  {\path{doi:10.1103/PhysRev.55.374}}.

\bibitem{Tolos2015}
L.~Tolos, J.~Schaffner-Bielich,
  \href{https://link.aps.org/doi/10.1103/PhysRevD.92.123002}{Dark compact
  planets}, Phys. Rev. D 92 (2015) 123002.
\newblock \href {http://dx.doi.org/10.1103/PhysRevD.92.123002}
  {\path{doi:10.1103/PhysRevD.92.123002}}.
\newline\urlprefix\url{https://link.aps.org/doi/10.1103/PhysRevD.92.123002}

\bibitem{Goldman:1989nd}
I.~Goldman, S.~Nussinov, {Weakly Interacting Massive Particles and Neutron
  Stars}, Phys. Rev. D40 (1989) 3221--3230.
\newblock \href {http://dx.doi.org/10.1103/PhysRevD.40.3221}
  {\path{doi:10.1103/PhysRevD.40.3221}}.

\bibitem{Kouvaris:2011fi}
C.~Kouvaris, P.~Tinyakov, {Excluding Light Asymmetric Bosonic Dark Matter},
  Phys. Rev. Lett. 107 (2011) 091301.
\newblock \href {http://arxiv.org/abs/1104.0382} {\path{arXiv:1104.0382}},
  \href {http://dx.doi.org/10.1103/PhysRevLett.107.091301}
  {\path{doi:10.1103/PhysRevLett.107.091301}}.

\bibitem{Keane:2012yh}
E.~F. Keane, B.~W. Stappers, M.~Kramer, A.~G. Lyne, {On the origin of a
  highly-dispersed coherent radio burst}, Mon. Not. Roy. Astron. Soc. 425
  (2012) 71.
\newblock \href {http://arxiv.org/abs/1206.4135} {\path{arXiv:1206.4135}},
  \href {http://dx.doi.org/10.1111/j.1745-3933.2012.01306.x}
  {\path{doi:10.1111/j.1745-3933.2012.01306.x}}.

\bibitem{Thornton:2013iua}
D.~Thornton, et~al., {A Population of Fast Radio Bursts at Cosmological
  Distances}, Science 341~(6141) (2013) 53--56.
\newblock \href {http://arxiv.org/abs/1307.1628} {\path{arXiv:1307.1628}},
  \href {http://dx.doi.org/10.1126/science.1236789}
  {\path{doi:10.1126/science.1236789}}.

\bibitem{Burke-Spolaor:2014rqa}
S.~Burke-Spolaor, K.~W. Bannister, {The Galactic Position Dependence of Fast
  Radio Bursts and the Discovery of FRB011025}, Astrophys. J. 792~(1) (2014)
  19.
\newblock \href {http://arxiv.org/abs/1407.0400} {\path{arXiv:1407.0400}},
  \href {http://dx.doi.org/10.1088/0004-637X/792/1/19}
  {\path{doi:10.1088/0004-637X/792/1/19}}.

\bibitem{Spitler:2014fla}
L.~G. Spitler, et~al., {Fast Radio Burst Discovered in the Arecibo Pulsar ALFA
  Survey}, Astrophys. J. 790~(2) (2014) 101.
\newblock \href {http://arxiv.org/abs/1404.2934} {\path{arXiv:1404.2934}},
  \href {http://dx.doi.org/10.1088/0004-637X/790/2/101}
  {\path{doi:10.1088/0004-637X/790/2/101}}.

\bibitem{Fuller:2014rza}
J.~Fuller, C.~Ott, {Dark Matter-induced Collapse of Neutron Stars: A Possible
  Link Between Fast Radio Bursts and the Missing Pulsar Problem}, Mon. Not.
  Roy. Astron. Soc. 450~(1) (2015) L71--L75.
\newblock \href {http://arxiv.org/abs/1412.6119} {\path{arXiv:1412.6119}},
  \href {http://dx.doi.org/10.1093/mnrasl/slv049}
  {\path{doi:10.1093/mnrasl/slv049}}.

\bibitem{Kouvaris2008}
C.~{Kouvaris}, {WIMP annihilation and cooling of neutron stars}, \prd 77~(2)
  (2008) 023006.
\newblock \href {http://arxiv.org/abs/0708.2362} {\path{arXiv:0708.2362}},
  \href {http://dx.doi.org/10.1103/PhysRevD.77.023006}
  {\path{doi:10.1103/PhysRevD.77.023006}}.

\bibitem{Bertone2008}
G.~{Bertone}, M.~{Fairbairn}, {Compact stars as dark matter probes}, \prd
  77~(4) (2008) 043515.
\newblock \href {http://arxiv.org/abs/0709.1485} {\path{arXiv:0709.1485}},
  \href {http://dx.doi.org/10.1103/PhysRevD.77.043515}
  {\path{doi:10.1103/PhysRevD.77.043515}}.

\bibitem{Kouvaris2010}
C.~Kouvaris, P.~Tinyakov, {Can Neutron stars constrain Dark Matter?}, Phys.
  Rev. D82 (2010) 063531.
\newblock \href {http://arxiv.org/abs/1004.0586} {\path{arXiv:1004.0586}},
  \href {http://dx.doi.org/10.1103/PhysRevD.82.063531}
  {\path{doi:10.1103/PhysRevD.82.063531}}.

\bibitem{McCullough:2010ai}
M.~McCullough, M.~Fairbairn, {Capture of Inelastic Dark Matter in White
  Dwarves}, Phys. Rev. D81 (2010) 083520.
\newblock \href {http://arxiv.org/abs/1001.2737} {\path{arXiv:1001.2737}},
  \href {http://dx.doi.org/10.1103/PhysRevD.81.083520}
  {\path{doi:10.1103/PhysRevD.81.083520}}.

\bibitem{deLavallaz2010}
A.~{de Lavallaz}, M.~{Fairbairn}, {Neutron stars as dark matter probes}, \prd
  81~(12) (2010) 123521.
\newblock \href {http://arxiv.org/abs/1004.0629} {\path{arXiv:1004.0629}},
  \href {http://dx.doi.org/10.1103/PhysRevD.81.123521}
  {\path{doi:10.1103/PhysRevD.81.123521}}.

\bibitem{PerezGarcia:2011hh}
M.~A. Perez-Garcia, J.~Silk, {Dark matter seeding and the kinematics and
  rotation of neutron stars}, Phys. Lett. B711 (2012) 6--9.
\newblock \href {http://arxiv.org/abs/1111.2275} {\path{arXiv:1111.2275}},
  \href {http://dx.doi.org/10.1016/j.physletb.2012.03.065}
  {\path{doi:10.1016/j.physletb.2012.03.065}}.

\bibitem{PerezGarcia:2010ap}
M.~A. Perez-Garcia, J.~Silk, J.~R. Stone, {Dark matter, neutron stars and
  strange quark matter}, Phys. Rev. Lett. 105 (2010) 141101.
\newblock \href {http://arxiv.org/abs/1007.1421} {\path{arXiv:1007.1421}},
  \href {http://dx.doi.org/10.1103/PhysRevLett.105.141101}
  {\path{doi:10.1103/PhysRevLett.105.141101}}.

\bibitem{Sandin2009}
F.~Sandin, P.~Ciarcelluti,
  \href{http://www.sciencedirect.com/science/article/pii/S0927650509001406}{Effects
  of mirror dark matter on neutron stars}, Astroparticle Physics 32~(5) (2009)
  278 -- 284.
\newblock \href
  {http://dx.doi.org/https://doi.org/10.1016/j.astropartphys.2009.09.005}
  {\path{doi:https://doi.org/10.1016/j.astropartphys.2009.09.005}}.
\newline\urlprefix\url{http://www.sciencedirect.com/science/article/pii/S0927650509001406}

\bibitem{Petraki2013}
K.~{Petraki}, R.~R. {Volkas}, {Review of Asymmetric Dark Matter}, International
  Journal of Modern Physics A 28 (2013) 1330028.
\newblock \href {http://arxiv.org/abs/1305.4939} {\path{arXiv:1305.4939}},
  \href {http://dx.doi.org/10.1142/S0217751X13300287}
  {\path{doi:10.1142/S0217751X13300287}}.

\bibitem{Ciarcelluti2010}
P.~Ciarcelluti, F.~Sandin,
  \href{http://www.sciencedirect.com/science/article/pii/S0370269310012955}{Have
  neutron stars a dark matter core?}, Physics Letters B 695~(1) (2011) 19 --
  21.
\newblock \href
  {http://dx.doi.org/https://doi.org/10.1016/j.physletb.2010.11.021}
  {\path{doi:https://doi.org/10.1016/j.physletb.2010.11.021}}.
\newline\urlprefix\url{http://www.sciencedirect.com/science/article/pii/S0370269310012955}

\bibitem{Kouvaris2013}
C.~{Kouvaris}, {The Dark Side of Neutron Stars}, ArXiv e-prints\href
  {http://arxiv.org/abs/1308.3222} {\path{arXiv:1308.3222}}.

\bibitem{Ciarcelluti:2010ji}
P.~Ciarcelluti, F.~Sandin, {Have neutron stars a dark matter core?}, Phys.
  Lett. B695 (2011) 19--21.
\newblock \href {http://arxiv.org/abs/1005.0857} {\path{arXiv:1005.0857}},
  \href {http://dx.doi.org/10.1016/j.physletb.2010.11.021}
  {\path{doi:10.1016/j.physletb.2010.11.021}}.

\bibitem{demorest2010}
P.~B. {Demorest}, T.~{Pennucci}, S.~M. {Ransom}, M.~S.~E. {Roberts}, J.~W.~T.
  {Hessels}, {A two-solar-mass neutron star measured using Shapiro delay}, \nat
  467 (2010) 1081--1083.
\newblock \href {http://arxiv.org/abs/1010.5788} {\path{arXiv:1010.5788}},
  \href {http://dx.doi.org/10.1038/nature09466}
  {\path{doi:10.1038/nature09466}}.

\bibitem{Antoniadis:2013pzd}
J.~Antoniadis, et~al., {A Massive Pulsar in a Compact Relativistic Binary},
  Science 340 (2013) 6131.
\newblock \href {http://arxiv.org/abs/1304.6875} {\path{arXiv:1304.6875}},
  \href {http://dx.doi.org/10.1126/science.1233232}
  {\path{doi:10.1126/science.1233232}}.

\bibitem{Rhoades:1974fn}
C.~E. Rhoades, Jr., R.~Ruffini, {Maximum mass of a neutron star}, Phys. Rev.
  Lett. 32 (1974) 324--327.
\newblock \href {http://dx.doi.org/10.1103/PhysRevLett.32.324}
  {\path{doi:10.1103/PhysRevLett.32.324}}.

\bibitem{Brecher:1976zz}
K.~Brecher, G.~Caporaso, {Neutron Stars within the Laws of Physics}, Annals N.
  Y. Acad. Sci. 302 (1977) 471--481.
\newblock \href {http://dx.doi.org/10.1111/j.1749-6632.1977.tb37067.x}
  {\path{doi:10.1111/j.1749-6632.1977.tb37067.x}}.

\bibitem{Margalit:2017dij}
B.~Margalit, B.~D. Metzger, {Constraining the Maximum Mass of Neutron Stars
  From Multi-Messenger Observations of GW170817}, Astrophys. J. 850~(2) (2017)
  L19.
\newblock \href {http://arxiv.org/abs/1710.05938} {\path{arXiv:1710.05938}},
  \href {http://dx.doi.org/10.3847/2041-8213/aa991c}
  {\path{doi:10.3847/2041-8213/aa991c}}.

\bibitem{Rezzolla2017}
L.~Rezzolla, E.~R. Most, L.~R. Weih, {Using gravitational-wave observations and
  quasi-universal relations to constrain the maximum mass of neutron stars},
  Astrophys. J. 852~(2) (2018) L25, [Astrophys. J. Lett.852,L25(2018)].
\newblock \href {http://arxiv.org/abs/1711.00314} {\path{arXiv:1711.00314}},
  \href {http://dx.doi.org/10.3847/2041-8213/aaa401}
  {\path{doi:10.3847/2041-8213/aaa401}}.

\bibitem{Stovall:2014gua}
K.~Stovall, et~al., {The Green Bank Northern Celestial Cap Pulsar Survey - I:
  Survey Description, Data Analysis, and Initial Results}, Astrophys. J.
  791~(1) (2014) 67.
\newblock \href {http://arxiv.org/abs/1406.5214} {\path{arXiv:1406.5214}},
  \href {http://dx.doi.org/10.1088/0004-637X/791/1/67}
  {\path{doi:10.1088/0004-637X/791/1/67}}.

\bibitem{Lynch:2018zxo}
R.~S. Lynch, et~al., {The Green Bank North Celestial Cap Pulsar Survey III: 45
  New Pulsar Timing Solutions}, Astrophys. J. 859~(2) (2018) 93.
\newblock \href {http://arxiv.org/abs/1805.04951} {\path{arXiv:1805.04951}},
  \href {http://dx.doi.org/10.3847/1538-4357/aabf8a}
  {\path{doi:10.3847/1538-4357/aabf8a}}.

\bibitem{Lattimer2004}
J.~M. Lattimer, M.~Prakash, {The physics of neutron stars}, Science 304 (2004)
  536--542.
\newblock \href {http://arxiv.org/abs/astro-ph/0405262}
  {\path{arXiv:astro-ph/0405262}}, \href
  {http://dx.doi.org/10.1126/science.1090720}
  {\path{doi:10.1126/science.1090720}}.

\bibitem{Li:2012ii}
A.~Li, F.~Huang, R.-X. Xu, {Too massive neutron stars: The role of dark
  matter?}, Astropart. Phys. 37 (2012) 70--74.
\newblock \href {http://arxiv.org/abs/1208.3722} {\path{arXiv:1208.3722}},
  \href {http://dx.doi.org/10.1016/j.astropartphys.2012.07.006}
  {\path{doi:10.1016/j.astropartphys.2012.07.006}}.

\bibitem{Sandin:2008db}
F.~Sandin, P.~Ciarcelluti, {Effects of mirror dark matter on neutron stars},
  Astropart. Phys. 32 (2009) 278--284.
\newblock \href {http://arxiv.org/abs/0809.2942} {\path{arXiv:0809.2942}},
  \href {http://dx.doi.org/10.1016/j.astropartphys.2009.09.005}
  {\path{doi:10.1016/j.astropartphys.2009.09.005}}.

\bibitem{Leung:2011zz}
S.~C. Leung, M.~C. Chu, L.~M. Lin, {Dark-matter admixed neutron stars}, Phys.
  Rev. D84 (2011) 107301.
\newblock \href {http://arxiv.org/abs/1111.1787} {\path{arXiv:1111.1787}},
  \href {http://dx.doi.org/10.1103/PhysRevD.84.107301}
  {\path{doi:10.1103/PhysRevD.84.107301}}.

\bibitem{Xiang:2013xwa}
Q.-F. Xiang, W.-Z. Jiang, D.-R. Zhang, R.-Y. Yang, {Effects of fermionic dark
  matter on properties of neutron stars}, Phys. Rev. C89~(2) (2014) 025803.
\newblock \href {http://arxiv.org/abs/1305.7354} {\path{arXiv:1305.7354}},
  \href {http://dx.doi.org/10.1103/PhysRevC.89.025803}
  {\path{doi:10.1103/PhysRevC.89.025803}}.

\bibitem{Goldman:2013qla}
I.~Goldman, R.~N. Mohapatra, S.~Nussinov, D.~Rosenbaum, V.~Teplitz, {Possible
  Implications of Asymmetric Fermionic Dark Matter for Neutron Stars}, Phys.
  Lett. B725 (2013) 200--207.
\newblock \href {http://arxiv.org/abs/1305.6908} {\path{arXiv:1305.6908}},
  \href {http://dx.doi.org/10.1016/j.physletb.2013.07.017}
  {\path{doi:10.1016/j.physletb.2013.07.017}}.

\bibitem{Leung2011}
S.-C. Leung, M.-C. Chu, L.-M. Lin,
  \href{https://link.aps.org/doi/10.1103/PhysRevD.84.107301}{Dark-matter
  admixed neutron stars}, Phys. Rev. D 84 (2011) 107301.
\newblock \href {http://dx.doi.org/10.1103/PhysRevD.84.107301}
  {\path{doi:10.1103/PhysRevD.84.107301}}.
\newline\urlprefix\url{https://link.aps.org/doi/10.1103/PhysRevD.84.107301}

\bibitem{Li2012}
A.~Li, F.~Huang, R.-X. Xu,
  \href{http://www.sciencedirect.com/science/article/pii/S0927650512001429}{Too
  massive neutron stars: The role of dark matter?}, Astroparticle Physics 37
  (2012) 70 -- 74.
\newblock \href
  {http://dx.doi.org/https://doi.org/10.1016/j.astropartphys.2012.07.006}
  {\path{doi:https://doi.org/10.1016/j.astropartphys.2012.07.006}}.
\newline\urlprefix\url{http://www.sciencedirect.com/science/article/pii/S0927650512001429}

\bibitem{Deliyergiyev2019}
M.~{Deliyergiyev}, A.~{Del Popolo}, L.~{Tolos}, M.~{Le Delliou}, X.~{Lee},
  F.~{Burgio}, {Dark compact objects: An extensive overview}, Phys. Rev.
  D99~(6) (2019) 063015.
\newblock \href {http://arxiv.org/abs/1903.01183} {\path{arXiv:1903.01183}},
  \href {http://dx.doi.org/10.1103/PhysRevD.99.063015}
  {\path{doi:10.1103/PhysRevD.99.063015}}.

\bibitem{Yang2011}
R.-J. {Yang}, X.-T. {Gao}, {Phase-space analysis of a class of k-essence
  cosmology}, Classical and Quantum Gravity 28~(6) (2011) 065012.
\newblock \href {http://arxiv.org/abs/1006.4986} {\path{arXiv:1006.4986}},
  \href {http://dx.doi.org/10.1088/0264-9381/28/6/065012}
  {\path{doi:10.1088/0264-9381/28/6/065012}}.

\bibitem{Kouvaris2011}
C.~{Kouvaris}, P.~{Tinyakov}, {Constraining asymmetric dark matter through
  observations of compact stars}, \prd 83~(8) (2011) 083512.
\newblock \href {http://arxiv.org/abs/1012.2039} {\path{arXiv:1012.2039}},
  \href {http://dx.doi.org/10.1103/PhysRevD.83.083512}
  {\path{doi:10.1103/PhysRevD.83.083512}}.

\bibitem{Guver2014}
T.~{G{\"u}ver}, A.~{Emre Erkoca}, M.~{Hall Reno}, I.~{Sarcevic}, {On the
  capture of dark matter by neutron stars}, \jcap 5 (2014) 013.
\newblock \href {http://arxiv.org/abs/1201.2400} {\path{arXiv:1201.2400}},
  \href {http://dx.doi.org/10.1088/1475-7516/2014/05/013}
  {\path{doi:10.1088/1475-7516/2014/05/013}}.

\bibitem{Zheng2016}
H.~{Zheng}, L.-W. {Chen}, {Strange Quark Stars as a Probe of Dark Matter}, \apj
  831 (2016) 127.
\newblock \href {http://arxiv.org/abs/1603.07518} {\path{arXiv:1603.07518}},
  \href {http://dx.doi.org/10.3847/0004-637X/831/2/127}
  {\path{doi:10.3847/0004-637X/831/2/127}}.

\bibitem{Aprile2017}
E.~{Aprile}, J.~{Aalbers}, F.~{Agostini}, M.~{Alfonsi}, F.~D. {Amaro},
  M.~{Anthony}, F.~{Arneodo}, P.~{Barrow}, L.~{Baudis}, B.~{Bauermeister},
  {First Dark Matter Search Results from the XENON1T Experiment}, \prl 119~(18)
  (2017) 181301.
\newblock \href {http://arxiv.org/abs/1705.06655} {\path{arXiv:1705.06655}},
  \href {http://dx.doi.org/10.1103/PhysRevLett.119.181301}
  {\path{doi:10.1103/PhysRevLett.119.181301}}.

\bibitem{Wong2010}
T.-W. {Wong}, B.~{Willems}, V.~{Kalogera}, {Constraints on Natal Kicks in
  Galactic Double Neutron Star Systems}, \apj 721~(2) (2010) 1689--1701.
\newblock \href {http://arxiv.org/abs/1008.2397} {\path{arXiv:1008.2397}},
  \href {http://dx.doi.org/10.1088/0004-637X/721/2/1689}
  {\path{doi:10.1088/0004-637X/721/2/1689}}.

\bibitem{Lorimer1998}
D.~R. {Lorimer}, {Binary and Millisecond Pulsars}, Living Reviews in Relativity
  1 (1998) 10.
\newblock \href {http://dx.doi.org/10.12942/lrr-1998-10}
  {\path{doi:10.12942/lrr-1998-10}}.

\bibitem{Taani:2012xp}
A.~Taani, L.~Naso, Y.~Wei, C.~Zhang, Y.~Zhao, {Modeling the Spatial
  Distribution of Neutron Stars in the Galaxy}, Astrophys. Space Sci. 341
  (2012) 601--609.
\newblock \href {http://arxiv.org/abs/1205.4307} {\path{arXiv:1205.4307}},
  \href {http://dx.doi.org/10.1007/s10509-012-1121-7}
  {\path{doi:10.1007/s10509-012-1121-7}}.

\bibitem{Ofek:2009wt}
E.~O. Ofek, {Space and velocity distributions of Galactic isolated old Neutron
  stars}, Publ. Astron. Soc. Pac. 121 (2009) 814.
\newblock \href {http://arxiv.org/abs/0910.3684} {\path{arXiv:0910.3684}},
  \href {http://dx.doi.org/10.1086/605389} {\path{doi:10.1086/605389}}.

\bibitem{Wei:2010}
Y.~Wei, C.~Zhang, X.~Wu, Y.~Zhao, C.~Chou, A.~Luo, {The spatial distribution of
  old neutron stars in the Galaxy}, Sci. China Phys. Mech. Astron. 53 (2010)
  1939–1946.
\newblock \href {http://dx.doi.org/10.1007/s11433-010-4104-0}
  {\path{doi:10.1007/s11433-010-4104-0}}.

\bibitem{Sartore:2009wn}
N.~Sartore, E.~Ripamonti, A.~Treves, R.~Turolla, {Galactic neutron stars I.
  Space and velocity distributions in the disk and in the halo}, Astron.
  Astrophys. 510 (2010) A23.
\newblock \href {http://arxiv.org/abs/0908.3182} {\path{arXiv:0908.3182}},
  \href {http://dx.doi.org/10.1051/0004-6361/200912222}
  {\path{doi:10.1051/0004-6361/200912222}}.

\bibitem{Dragicevich:1999}
P.~M. {Dragicevich}, D.~G. {Blair}, R.~R. {Burman}, {Why are supernovae in our
  Galaxy so frequent?}, \mnras 302 (1999) 693--699.
\newblock \href {http://dx.doi.org/10.1046/j.1365-8711.1999.02145.x}
  {\path{doi:10.1046/j.1365-8711.1999.02145.x}}.

\bibitem{Zhong2012}
Y.-z. {Fan}, R.-z. {Yang}, J.~{Chang}, {Constraining Asymmetric Bosonic
  Non-interacting Dark Matter with Neutron Stars}, ArXiv e-prints\href
  {http://arxiv.org/abs/1204.2564} {\path{arXiv:1204.2564}}.

\bibitem{Berezinsky2013}
V.~S. {Berezinsky}, V.~I. {Dokuchaev}, Y.~N. {Eroshenko}, {Formation and
  internal structure of superdense dark matter clumps and ultracompact
  minihaloes}, \jcap 11 (2013) 059.
\newblock \href {http://arxiv.org/abs/1308.6742} {\path{arXiv:1308.6742}},
  \href {http://dx.doi.org/10.1088/1475-7516/2013/11/059}
  {\path{doi:10.1088/1475-7516/2013/11/059}}.

\bibitem{Chang2018}
J.~H. Chang, D.~Egana-Ugrinovic, R.~Essig, C.~Kouvaris, {Structure Formation
  and Exotic Compact Objects in a Dissipative Dark Sector}\href
  {http://arxiv.org/abs/1812.07000} {\path{arXiv:1812.07000}}.

\bibitem{Ullio2010}
R.~Catena, P.~Ullio, {A novel determination of the local dark matter density},
  JCAP 1008 (2010) 004.
\newblock \href {http://arxiv.org/abs/0907.0018} {\path{arXiv:0907.0018}},
  \href {http://dx.doi.org/10.1088/1475-7516/2010/08/004}
  {\path{doi:10.1088/1475-7516/2010/08/004}}.

\bibitem{Weber2010}
M.~Weber, W.~de~Boer, {Determination of the Local Dark Matter Density in our
  Galaxy}, Astron. Astrophys. 509 (2010) A25.
\newblock \href {http://arxiv.org/abs/0910.4272} {\path{arXiv:0910.4272}},
  \href {http://dx.doi.org/10.1051/0004-6361/200913381}
  {\path{doi:10.1051/0004-6361/200913381}}.

\bibitem{Navarro1997}
J.~F. {Navarro}, C.~S. {Frenk}, S.~D.~M. {White}, {A Universal Density Profile
  from Hierarchical Clustering}, \apj 490 (1997) 493--+.
\newblock \href {http://arxiv.org/abs/astro-ph/9611107}
  {\path{arXiv:astro-ph/9611107}}, \href {http://dx.doi.org/10.1086/304888}
  {\path{doi:10.1086/304888}}.

\bibitem{Stadel2009}
J.~{Stadel}, D.~{Potter}, B.~{Moore}, J.~{Diemand}, P.~{Madau}, M.~{Zemp},
  M.~{Kuhlen}, V.~{Quilis}, {Quantifying the heart of darkness with GHALO - a
  multibillion particle simulation of a galactic halo}, \mnras 398 (2009)
  L21--L25.
\newblock \href {http://arxiv.org/abs/0808.2981} {\path{arXiv:0808.2981}},
  \href {http://dx.doi.org/10.1111/j.1745-3933.2009.00699.x}
  {\path{doi:10.1111/j.1745-3933.2009.00699.x}}.

\bibitem{Navarro2010}
J.~F. {Navarro}, A.~{Ludlow}, V.~{Springel}, J.~{Wang}, M.~{Vogelsberger},
  S.~D.~M. {White}, A.~{Jenkins}, C.~S. {Frenk}, A.~{Helmi}, {The diversity and
  similarity of simulated cold dark matter haloes}, \mnras 402 (2010) 21--34.
\newblock \href {http://arxiv.org/abs/0810.1522} {\path{arXiv:0810.1522}},
  \href {http://dx.doi.org/10.1111/j.1365-2966.2009.15878.x}
  {\path{doi:10.1111/j.1365-2966.2009.15878.x}}.

\bibitem{DiCintio2014}
A.~Di~Cintio, C.~B. Brook, A.~A. Dutton, A.~V. Macciò, G.~S. Stinson,
  A.~Knebe, {A mass-dependent density profile for dark matter haloes including
  the influence of galaxy formation}, Mon. Not. Roy. Astron. Soc. 441~(4)
  (2014) 2986--2995.
\newblock \href {http://arxiv.org/abs/1404.5959} {\path{arXiv:1404.5959}},
  \href {http://dx.doi.org/10.1093/mnras/stu729}
  {\path{doi:10.1093/mnras/stu729}}.

\bibitem{Iocco2011}
F.~Iocco, M.~Pato, G.~Bertone, P.~Jetzer, {Dark Matter distribution in the
  Milky Way: microlensing and dynamical constraints}, JCAP 1111 (2011) 029.
\newblock \href {http://arxiv.org/abs/1107.5810} {\path{arXiv:1107.5810}},
  \href {http://dx.doi.org/10.1088/1475-7516/2011/11/029}
  {\path{doi:10.1088/1475-7516/2011/11/029}}.

\bibitem{Udrescu2018}
S.~M. Udrescu, A.~A. Dutton, A.~V. Macci\`o, T.~Buck, {A deeper look into the
  structure of {$\Lambda$}CDM haloes: correlations between halo parameters from
  Einasto fits}, Mon. Not. Roy. Astron. Soc. 482 (2019) 5259.
\newblock \href {http://arxiv.org/abs/1811.04955} {\path{arXiv:1811.04955}},
  \href {http://dx.doi.org/10.1093/mnras/sty3112}
  {\path{doi:10.1093/mnras/sty3112}}.

\bibitem{Dutton2014}
A.~A. Dutton, A.~V. Macciò, {Cold dark matter haloes in the Planck era:
  evolution of structural parameters for Einasto and NFW profiles}, Mon. Not.
  Roy. Astron. Soc. 441~(4) (2014) 3359--3374.
\newblock \href {http://arxiv.org/abs/1402.7073} {\path{arXiv:1402.7073}},
  \href {http://dx.doi.org/10.1093/mnras/stu742}
  {\path{doi:10.1093/mnras/stu742}}.

\bibitem{Gnedin2004}
O.~Y. {Gnedin}, A.~V. {Kravtsov}, A.~A. {Klypin}, D.~{Nagai}, {Response of Dark
  Matter Halos to Condensation of Baryons: Cosmological Simulations and
  Improved Adiabatic Contraction Model}, \apj 616 (2004) 16--26.
\newblock \href {http://arxiv.org/abs/astro-ph/0406247}
  {\path{arXiv:astro-ph/0406247}}, \href {http://dx.doi.org/10.1086/424914}
  {\path{doi:10.1086/424914}}.

\bibitem{Gustafsson2006}
M.~{Gustafsson}, M.~{Fairbairn}, J.~{Sommer-Larsen}, {Baryonic pinching of
  galactic dark matter halos}, \prd 74~(12) (2006) 123522.
\newblock \href {http://arxiv.org/abs/astro-ph/0608634}
  {\path{arXiv:astro-ph/0608634}}, \href
  {http://dx.doi.org/10.1103/PhysRevD.74.123522}
  {\path{doi:10.1103/PhysRevD.74.123522}}.

\bibitem{Pedrosa2009}
S.~E. Pedrosa, P.~B. Tissera, C.~Scannapieco, {The impact of baryons on dark
  matter haloes}, Mon. Not. Roy. Astron. Soc. 395 (2009) 57.
\newblock \href {http://arxiv.org/abs/0902.2100} {\path{arXiv:0902.2100}},
  \href {http://dx.doi.org/10.1111/j.1745-3933.2009.00641.x}
  {\path{doi:10.1111/j.1745-3933.2009.00641.x}}.

\bibitem{Duffy2010}
A.~R. Duffy, J.~Schaye, S.~T. Kay, C.~Dalla~Vecchia, R.~A. Battye, C.~M. Booth,
  {Impact of baryon physics on dark matter structures: a detailed simulation
  study of halo density profiles}, Mon. Not. Roy. Astron. Soc. 405 (2010) 2161.
\newblock \href {http://arxiv.org/abs/1001.3447} {\path{arXiv:1001.3447}},
  \href {http://dx.doi.org/10.1111/j.1365-2966.2010.16613.x}
  {\path{doi:10.1111/j.1365-2966.2010.16613.x}}.

\bibitem{DelPopolo2010}
A.~{Del Popolo}, {On the universality of density profiles}, \mnras 408 (2010)
  1808--1817.
\newblock \href {http://arxiv.org/abs/1012.4322} {\path{arXiv:1012.4322}},
  \href {http://dx.doi.org/10.1111/j.1365-2966.2010.17288.x}
  {\path{doi:10.1111/j.1365-2966.2010.17288.x}}.

\bibitem{DelPopolo2016a}
A.~{Del Popolo}, F.~{Pace}, {The Cusp/Core problem: supernovae feedback versus
  the baryonic clumps and dynamical friction model}, \apss 361 (2016) 162.
\newblock \href {http://arxiv.org/abs/1502.01947} {\path{arXiv:1502.01947}},
  \href {http://dx.doi.org/10.1007/s10509-016-2742-z}
  {\path{doi:10.1007/s10509-016-2742-z}}.

\bibitem{DelPopolo2017}
A.~{Del Popolo}, F.~{Pace}, M.~{Le Delliou}, {A high precision semi-analytic
  mass function}, \jcap 3 (2017) 032.
\newblock \href {http://arxiv.org/abs/1703.06918} {\path{arXiv:1703.06918}},
  \href {http://dx.doi.org/10.1088/1475-7516/2017/03/032}
  {\path{doi:10.1088/1475-7516/2017/03/032}}.

\bibitem{DelPopolo2009}
A.~{Del Popolo}, {The Cusp/Core Problem and the Secondary Infall Model}, \apj
  698 (2009) 2093--2113.
\newblock \href {http://arxiv.org/abs/0906.4447} {\path{arXiv:0906.4447}},
  \href {http://dx.doi.org/10.1088/0004-637X/698/2/2093}
  {\path{doi:10.1088/0004-637X/698/2/2093}}.

\bibitem{Prada2004}
F.~Prada, A.~Klypin, J.~Flix~Molina, M.~Martinez, E.~Simonneau, {Dark Matter
  Annihilation in the Milky Way Galaxy: Effects of Baryonic Compression}, Phys.
  Rev. Lett. 93 (2004) 241301.
\newblock \href {http://arxiv.org/abs/astro-ph/0401512}
  {\path{arXiv:astro-ph/0401512}}, \href
  {http://dx.doi.org/10.1103/PhysRevLett.93.241301}
  {\path{doi:10.1103/PhysRevLett.93.241301}}.

\bibitem{Cirelli2010}
M.~Cirelli, G.~Corcella, A.~Hektor, G.~Hutsi, M.~Kadastik, P.~Panci, M.~Raidal,
  F.~Sala, A.~Strumia, {PPPC 4 DM ID: A Poor Particle Physicist Cookbook for
  Dark Matter Indirect Detection}, JCAP 1103 (2011) 051, [Erratum:
  JCAP1210,E01(2012)].
\newblock \href {http://arxiv.org/abs/1012.4515} {\path{arXiv:1012.4515}},
  \href {http://dx.doi.org/10.1088/1475-7516/2012/10/E01,
  10.1088/1475-7516/2011/03/051} {\path{doi:10.1088/1475-7516/2012/10/E01,
  10.1088/1475-7516/2011/03/051}}.

\bibitem{Maccio2008}
A.~V. {Macci{\`o}}, A.~A. {Dutton}, F.~C. {van den Bosch}, {Concentration, spin
  and shape of dark matter haloes as a function of the cosmological model:
  WMAP1, WMAP3 and WMAP5 results}, \mnras 391 (2008) 1940--1954.
\newblock \href {http://arxiv.org/abs/0805.1926} {\path{arXiv:0805.1926}},
  \href {http://dx.doi.org/10.1111/j.1365-2966.2008.14029.x}
  {\path{doi:10.1111/j.1365-2966.2008.14029.x}}.

\bibitem{Duffy2008}
A.~R. {Duffy}, J.~{Schaye}, S.~T. {Kay}, C.~{Dalla Vecchia}, {Dark matter halo
  concentrations in the Wilkinson Microwave Anisotropy Probe year 5 cosmology},
  \mnras 390 (2008) L64--L68.
\newblock \href {http://arxiv.org/abs/0804.2486} {\path{arXiv:0804.2486}},
  \href {http://dx.doi.org/10.1111/j.1745-3933.2008.00537.x}
  {\path{doi:10.1111/j.1745-3933.2008.00537.x}}.

\bibitem{Deason2012}
A.~J. {Deason}, V.~{Belokurov}, N.~W. {Evans}, J.~{An}, {Broken degeneracies:
  the rotation curve and velocity anisotropy of the Milky Way halo}, \mnras 424
  (2012) L44--L48.
\newblock \href {http://arxiv.org/abs/1204.5189} {\path{arXiv:1204.5189}},
  \href {http://dx.doi.org/10.1111/j.1745-3933.2012.01283.x}
  {\path{doi:10.1111/j.1745-3933.2012.01283.x}}.

\bibitem{Nesti2013}
F.~Nesti, P.~Salucci, {The Dark Matter halo of the Milky Way, AD 2013}, JCAP
  1307 (2013) 016.
\newblock \href {http://arxiv.org/abs/1304.5127} {\path{arXiv:1304.5127}},
  \href {http://dx.doi.org/10.1088/1475-7516/2013/07/016}
  {\path{doi:10.1088/1475-7516/2013/07/016}}.

\bibitem{Pato2015}
M.~{Pato}, F.~{Iocco}, G.~{Bertone}, {Dynamical constraints on the dark matter
  distribution in the Milky Way}, \jcap 12 (2015) 001.
\newblock \href {http://arxiv.org/abs/1504.06324} {\path{arXiv:1504.06324}},
  \href {http://dx.doi.org/10.1088/1475-7516/2015/12/001}
  {\path{doi:10.1088/1475-7516/2015/12/001}}.

\bibitem{Xue2008}
X.~X. Xue, et~al., {The Milky Way's Circular Velocity Curve to 60 kpc and an
  Estimate of the Dark Matter Halo Mass from Kinematics of ~2400 SDSS Blue
  Horizontal Branch Stars}, Astrophys. J. 684 (2008) 1143--1158.
\newblock \href {http://arxiv.org/abs/0801.1232} {\path{arXiv:0801.1232}},
  \href {http://dx.doi.org/10.1086/589500} {\path{doi:10.1086/589500}}.

\bibitem{Bernal2012}
N.~Bernal, S.~Palomares-Ruiz, {Constraining the Milky Way Dark Matter Density
  Profile with Gamma-Rays with Fermi-LAT}, JCAP 1201 (2012) 006.
\newblock \href {http://arxiv.org/abs/1103.2377} {\path{arXiv:1103.2377}},
  \href {http://dx.doi.org/10.1088/1475-7516/2012/01/006}
  {\path{doi:10.1088/1475-7516/2012/01/006}}.

\bibitem{Catena2010}
R.~{Catena}, P.~{Ullio}, {A novel determination of the local dark matter
  density}, \jcap 8 (2010) 004.
\newblock \href {http://arxiv.org/abs/0907.0018} {\path{arXiv:0907.0018}},
  \href {http://dx.doi.org/10.1088/1475-7516/2010/08/004}
  {\path{doi:10.1088/1475-7516/2010/08/004}}.

\bibitem{Karukes2019}
E.~V. {Karukes}, M.~{Benito}, F.~{Iocco}, R.~{Trotta}, A.~{Geringer-Sameth},
  {Bayesian reconstruction of the Milky Way dark matter distribution}, arXiv
  e-prints\href {http://arxiv.org/abs/1901.02463} {\path{arXiv:1901.02463}}.

\bibitem{Ivanytskyi:2019wxd}
O.~Ivanytskyi, V.~Sagun, I.~Lopes, {Neutron stars: new constraints on
  asymmetric dark matter}\href {http://arxiv.org/abs/1910.09925}
  {\path{arXiv:1910.09925}}.

\bibitem{BertoneMerritt2005}
G.~{Bertone}, D.~{Merritt}, {Time-dependent models for dark matter at the
  galactic center}, \prd 72~(10) (2005) 103502.
\newblock \href {http://arxiv.org/abs/astro-ph/0501555}
  {\path{arXiv:astro-ph/0501555}}, \href
  {http://dx.doi.org/10.1103/PhysRevD.72.103502}
  {\path{doi:10.1103/PhysRevD.72.103502}}.

\bibitem{Gondolo1999}
P.~Gondolo, J.~Silk, {Dark matter annihilation at the galactic center}, Phys.
  Rev. Lett. 83 (1999) 1719--1722.
\newblock \href {http://arxiv.org/abs/astro-ph/9906391}
  {\path{arXiv:astro-ph/9906391}}, \href
  {http://dx.doi.org/10.1103/PhysRevLett.83.1719}
  {\path{doi:10.1103/PhysRevLett.83.1719}}.

\bibitem{Sandick2018}
P.~{Sandick}, T.~{Yamamoto}, K.~{Sinha}, {Black holes, dark matter spikes, and
  constraints on simplified models with t -channel mediators}, \prd 98~(3)
  (2018) 035004.
\newblock \href {http://arxiv.org/abs/1701.00067} {\path{arXiv:1701.00067}},
  \href {http://dx.doi.org/10.1103/PhysRevD.98.035004}
  {\path{doi:10.1103/PhysRevD.98.035004}}.

\bibitem{Lacroix2018}
T.~{Lacroix}, {Dynamical constraints on a dark matter spike at the Galactic
  centre from stellar orbits}, \aap 619 (2018) A46.
\newblock \href {http://arxiv.org/abs/1801.01308} {\path{arXiv:1801.01308}},
  \href {http://dx.doi.org/10.1051/0004-6361/201832652}
  {\path{doi:10.1051/0004-6361/201832652}}.

\bibitem{Bennewitz2019}
E.~R. {Bennewitz}, C.~{Gaidau}, T.~W. {Baumgarte}, S.~L. {Shapiro}, {Dark
  matter heating of gas accreting onto Sgr A$^*$}, arXiv e-prints (2019)
  arXiv:1907.00015\href {http://arxiv.org/abs/1907.00015}
  {\path{arXiv:1907.00015}}.

\bibitem{Fields2014}
B.~D. {Fields}, S.~L. {Shapiro}, J.~{Shelton}, {Galactic Center Gamma-Ray
  Excess from Dark Matter Annihilation: Is There a Black Hole Spike?}, \prl
  113~(15) (2014) 151302.
\newblock \href {http://arxiv.org/abs/1406.4856} {\path{arXiv:1406.4856}},
  \href {http://dx.doi.org/10.1103/PhysRevLett.113.151302}
  {\path{doi:10.1103/PhysRevLett.113.151302}}.

\bibitem{ShapiroShelton2016}
S.~L. {Shapiro}, J.~{Shelton}, {Weak annihilation cusp inside the dark matter
  spike about a black hole}, \prd 93~(12) (2016) 123510.
\newblock \href {http://arxiv.org/abs/1606.01248} {\path{arXiv:1606.01248}},
  \href {http://dx.doi.org/10.1103/PhysRevD.93.123510}
  {\path{doi:10.1103/PhysRevD.93.123510}}.

\bibitem{Iorio2010}
L.~Iorio, {Phenomenological constraints on accretion of non-annihilating dark
  matter on the PSR B1257+12 pulsar from orbital dynamics of its planets}, JCAP
  1011 (2010) 046.
\newblock \href {http://arxiv.org/abs/1005.5078} {\path{arXiv:1005.5078}},
  \href {http://dx.doi.org/10.1088/1475-7516/2010/11/046}
  {\path{doi:10.1088/1475-7516/2010/11/046}}.

\bibitem{DiCintio:2014xia}
A.~Di~Cintio, C.~B. Brook, A.~A. Dutton, A.~V. Macciò, G.~S. Stinson,
  A.~Knebe, {A mass-dependent density profile for dark matter haloes including
  the influence of galaxy formation}, Mon. Not. Roy. Astron. Soc. 441~(4)
  (2014) 2986--2995.
\newblock \href {http://arxiv.org/abs/1404.5959} {\path{arXiv:1404.5959}},
  \href {http://dx.doi.org/10.1093/mnras/stu729}
  {\path{doi:10.1093/mnras/stu729}}.

\bibitem{Pfahl2003}
E.~Pfahl, A.~Loeb, {Probing the spacetime around Sgr A* with radio pulsars},
  Astrophys. J. 615 (2004) 253--258.
\newblock \href {http://arxiv.org/abs/astro-ph/0309744}
  {\path{arXiv:astro-ph/0309744}}, \href {http://dx.doi.org/10.1086/423975}
  {\path{doi:10.1086/423975}}.

\bibitem{Wharton2011}
R.~S. Wharton, S.~Chatterjee, J.~M. Cordes, J.~S. Deneva, T.~J.~W. Lazio,
  {Multiwavelength Constraints on Pulsar Populations in the Galactic Center},
  Astrophys. J. 753 (2012) 108.
\newblock \href {http://arxiv.org/abs/1111.4216} {\path{arXiv:1111.4216}},
  \href {http://dx.doi.org/10.1088/0004-637X/753/2/108}
  {\path{doi:10.1088/0004-637X/753/2/108}}.

\bibitem{Chennamangalam2013}
J.~Chennamangalam, D.~R. Lorimer, {The Galactic centre pulsar population}, Mon.
  Not. Roy. Astron. Soc. 440 (2014) 86.
\newblock \href {http://arxiv.org/abs/1311.4846} {\path{arXiv:1311.4846}},
  \href {http://dx.doi.org/10.1093/mnrasl/slu025}
  {\path{doi:10.1093/mnrasl/slu025}}.

\bibitem{Eatough2013}
R.~P. Eatough, et~al., {A strong magnetic field around the supermassive black
  hole at the centre of the Galaxy}, Nature 501 (2013) 391--394.
\newblock \href {http://arxiv.org/abs/1308.3147} {\path{arXiv:1308.3147}},
  \href {http://dx.doi.org/10.1038/nature12499}
  {\path{doi:10.1038/nature12499}}.

\bibitem{Mori2013}
K.~Mori, et~al., {NuSTAR discovery of a 3.76-second transient magnetar near
  Sagittarius A*}, Astrophys. J. 770 (2013) L23.
\newblock \href {http://arxiv.org/abs/1305.1945} {\path{arXiv:1305.1945}},
  \href {http://dx.doi.org/10.1088/2041-8205/770/2/L23}
  {\path{doi:10.1088/2041-8205/770/2/L23}}.

\bibitem{Kennea2013}
J.~A. Kennea, et~al., {Swift Discovery of a New Soft Gamma Repeater, SGR
  J1745-29, near Sagittarius A*}, Astrophys. J. 770 (2013) L24.
\newblock \href {http://arxiv.org/abs/1305.2128} {\path{arXiv:1305.2128}},
  \href {http://dx.doi.org/10.1088/2041-8205/770/2/L24}
  {\path{doi:10.1088/2041-8205/770/2/L24}}.

\bibitem{Shannon2013}
R.~M. Shannon, S.~Johnston, {Radio properties of the magnetar near Sagittarius
  A* from observations with the Australia Telescope Compact Array}, Mon. Not.
  Roy. Astron. Soc. 435 (2013) 29.
\newblock \href {http://arxiv.org/abs/1305.3036} {\path{arXiv:1305.3036}},
  \href {http://dx.doi.org/10.1093/mnrasl/slt088}
  {\path{doi:10.1093/mnrasl/slt088}}.

\bibitem{Lazio1998}
T.~J.~W. Lazio, J.~M. Cordes, {Hyperstrong radio-wave scattering in the
  galactic center. 2. A likelihood analysis of free electrons in the galactic
  center}, Astrophys. J. 505 (1998) 715.
\newblock \href {http://arxiv.org/abs/astro-ph/9804157}
  {\path{arXiv:astro-ph/9804157}}, \href {http://dx.doi.org/10.1086/306174}
  {\path{doi:10.1086/306174}}.

\bibitem{Bower2018}
G.~C. Bower, et~al., {Galactic Center Pulsars with the ngVLA}\href
  {http://arxiv.org/abs/1810.06623} {\path{arXiv:1810.06623}}.

\bibitem{FaucherGiguere2010}
C.~A. Faucher-Giguere, A.~Loeb, {Pulsar-Black Hole Binaries in the Galactic
  Center}, Mon. Not. Roy. Astron. Soc. 415 (2011) 3951.
\newblock \href {http://arxiv.org/abs/1012.0573} {\path{arXiv:1012.0573}},
  \href {http://dx.doi.org/10.1111/j.1365-2966.2011.19019.x}
  {\path{doi:10.1111/j.1365-2966.2011.19019.x}}.

\bibitem{Rajwade2017}
K.~M. {Rajwade}, D.~R. {Lorimer}, L.~D. {Anderson}, {Detecting pulsars in the
  Galactic Centre}, \mnras 471 (2017) 730--739.
\newblock \href {http://arxiv.org/abs/1611.06977} {\path{arXiv:1611.06977}},
  \href {http://dx.doi.org/10.1093/mnras/stx1661}
  {\path{doi:10.1093/mnras/stx1661}}.

\bibitem{Murphy:2018vxa}
E.~J. Murphy, et~al., {Science with an ngVLA: The ngVLA Science Case and
  Associated Science Requirements}, ASP Conf. Ser. 517 (2018) 3.
\newblock \href {http://arxiv.org/abs/1810.07524} {\path{arXiv:1810.07524}}.

\bibitem{Keane2015}
E.~{Keane}, B.~{Bhattacharyya}, M.~{Kramer}, B.~{Stappers}, E.~F. {Keane},
  B.~{Bhattacharyya}, M.~{Kramer}, B.~W. {Stappers}, S.~D. {Bates},
  M.~{Burgay}, {A Cosmic Census of Radio Pulsars with the SKA}, in: Advancing
  Astrophysics with the Square Kilometre Array (AASKA14), 2015, p.~40.
\newblock \href {http://arxiv.org/abs/1501.00056} {\path{arXiv:1501.00056}}.

\bibitem{Luminet2019}
J.-P. {Luminet}, {An Illustrated History of Black Hole Imaging : Personal
  Recollections (1972-2002)}, arXiv e-prints\href
  {http://arxiv.org/abs/1902.11196} {\path{arXiv:1902.11196}}.

\bibitem{Engineer1998}
S.~Engineer, K.~Srinivasan, T.~Padmanabhan, {A Formal analysis of
  two-dimensional gravity}, Astrophys. J. 512 (1999) 1.
\newblock \href {http://arxiv.org/abs/astro-ph/9805192}
  {\path{arXiv:astro-ph/9805192}}, \href {http://dx.doi.org/10.1086/306753}
  {\path{doi:10.1086/306753}}.

\bibitem{Watts:2016uzu}
A.~L. Watts, et~al., {Colloquium : Measuring the neutron star equation of state
  using x-ray timing}, Rev. Mod. Phys. 88~(2) (2016) 021001.
\newblock \href {http://arxiv.org/abs/1602.01081} {\path{arXiv:1602.01081}},
  \href {http://dx.doi.org/10.1103/RevModPhys.88.021001}
  {\path{doi:10.1103/RevModPhys.88.021001}}.

\bibitem{Ho:2015vza}
W.~C.~G. Ho, C.~M. Espinoza, D.~Antonopoulou, N.~Andersson, {Pinning down the
  superfluid and measuring masses using pulsar glitches}\href
  {http://arxiv.org/abs/1510.00395} {\path{arXiv:1510.00395}}, \href
  {http://dx.doi.org/10.1126/sciadv.1500578}
  {\path{doi:10.1126/sciadv.1500578}}.

\bibitem{Konar:2016lgc}
S.~Konar, et~al., {Neutron Star Physics in the Square Kilometer Array Era : An
  Indian Perspective}, J. Astrophys. Astron. 37 (2016) 36.
\newblock \href {http://arxiv.org/abs/1610.08175} {\path{arXiv:1610.08175}},
  \href {http://dx.doi.org/10.1007/s12036-016-9409-6}
  {\path{doi:10.1007/s12036-016-9409-6}}.

\bibitem{Barcons:2012zb}
X.~Barcons, et~al., {Athena (Advanced Telescope for High ENergy Astrophysics)
  Assessment Study Report for ESA Cosmic Vision 2015-2025}\href
  {http://arxiv.org/abs/1207.2745} {\path{arXiv:1207.2745}}.

\bibitem{Athena:2014cdf}
ESA, {ATHENA. Assessment of an X-Ray Telescope for the ESA Cosmic Vision
  Program}, CDF Study Report CDF-150(A) (2014) 332.

\bibitem{NICER:2012}
K.~C. {Gendreau}, Z.~{Arzoumanian}, T.~{Okajima}, {The Neutron star Interior
  Composition ExploreR (NICER): an Explorer mission of opportunity for soft
  x-ray timing spectroscopy}, in: Society of Photo-Optical Instrumentation
  Engineers (SPIE) Conference Series, Vol. 8443 of Society of Photo-Optical
  Instrumentation Engineers (SPIE) Conference Series, 2012.
\newblock \href {http://dx.doi.org/10.1117/12.926396}
  {\path{doi:10.1117/12.926396}}.

\bibitem{Watts:2018iom}
A.~L. Watts, et~al., {Dense matter with eXTP}, Sci. China Phys. Mech. Astron.
  62~(2) (2019) 29503.
\newblock \href {http://dx.doi.org/10.1007/s11433-017-9188-4}
  {\path{doi:10.1007/s11433-017-9188-4}}.

\end{thebibliography}

\end{document}